\documentclass[a4paper,10pt]{article}
\usepackage[a4paper, total={6in, 8in}]{geometry}

\usepackage{graphicx,multicol,subeqnarray,longtable,amsmath}

\newcommand{\vol}[1]{\textbf{#1}}
\newcommand{\journal}[1]{\emph{#1}}

\begin{document}
%






\title{UPDATED RADIO $\Sigma-D$ RELATION
FOR GALACTIC SUPERNOVA REMNANTS -- II}

\author{B.~Vukoti\'c$^1$\\
{\small \tt bvukotic@aob.rs}
\and
A.~{\'C}iprijanovi\'c $^{2,3}$\\
{\small \tt  aleksandra@matf.bg.ac.rs}
\and
M.~M.~Vu\v ceti\'c$^2$\\
{\small \tt mandjelic@matf.bg.ac.rs}
\and
D.~Oni\'c$^2$\\
{\small \tt donic@matf.bg.ac.rs}
\and
D.~Uro\v sevi\'c$^2$\\
{\small \tt dejanu@math.rs}
\and
\and
$^1$\emph{\small Astronomical Observatory Volgina 7, 11060 Belgrade 38, Serbia}\\
$^2$\emph{\small Department of Astronomy, Faculty of Mathematics, University of Belgrade}\\
\emph{\small Studentski trg 16, 11000 Belgrade, Serbia}\\
$^3$\emph{\small Mathematical Institute of the Serbian Academy of Sciences and Arts}\\
\emph{\small Kneza Mihaila 36, 11000 Belgrade, Serbia}
}

\date{}

\maketitle



\begin{abstract} In this paper we present the updated empirical radio surface-brightness-to-diameter ($\Sigma$--$D$) relation for Galactic supernova remnants (SNRs) calibrated using $110$ SNRs with reliable distances. We apply orthogonal fitting procedure and kernel density smoothing in $\Sigma-D$ plane and compare the results with the latest theoretical $\Sigma-D$ relations derived from simulations of radio evolution of SNRs. We argue that the best agreement between the empirical and simulated $\Sigma-D$ relations is achieved if the mixed-morphology SNRs and SNRs of both, low brightness and small diameter, are filtered out from the calibration sample. The distances to $5$ newly discovered remnants and $27$ new candidates for shell SNRs are estimated from our full and filtered calibration samples. 
\end{abstract}



\setlength{\tabcolsep}{1pt}
\section{Introduction}

Studying Galactic supernova remnants (SNRs) and their properties is crucial for understanding the evolution of interstellar matter in the universe. Still, studying even the most basic properties of SNRs, such as their distance, is quite difficult. Galactic SNRs are mostly observed in the radio domain, and the updated version of Galactic SNR catalog (Green 
2019) lists $294$ objects, where only a part of the observed sample actually has their distance determined from some independent empirical method (Green 1984). Some of the methods for SNR distance determination include: coincidences with the observed \hbox{H\,{\sc ii}} regions and molecular clouds (Eger et al.~2011, Ranasinghe and Leahy 2018a, 2018b, Supan et al.~2018, Yu et al.~2019), \hbox{H\,{\sc i}} absorption features and polarization (Yar-Uyanıker et al.~2004, Ranasinghe and Leahy~2018a, 2018b), association with the red clump stars (Shan et al.~2018, 2019), pulsars (Chatterjee et al.~2009), the optical proper motion and H$\alpha$ line radial velocity measurements (Lozinskaya et al.~1993, Katsuda et al.~2016), etc. When distance determination is not possible by using some of above mentioned methods, some estimates can be inferred using the radio-surface-brightness-to-diameter ($\Sigma$ -- $D$) relation, first proposed by Shklovsky (1960):
\begin{equation}
\Sigma_\nu (D) = AD^{\beta},
\end{equation}
where $A$ depends on properties of the supernova explosion, and exponent $\beta$ depends on the spectral index $\alpha$ of the integrated radio emission from the remnant $S_\nu \propto \nu^{-\alpha}$, where $S_\nu$ is the flux density at frequency $\nu$. The theoretically derived $\Sigma$--$D$ relations have slopes $\beta$ between $-3$ and $-6$ (e.g. Uro{\v s}evi{\' c} (2005, and reference therein), Berezhko and V{\"o}lk 2004, Pavlovi\'c et al.~(2013, hereafter Paper I), Pavlovi\'c et al.~2018). Calibration of the $\Sigma-D$ relation is usually performed on a linear version of Eq. (1):
\begin{equation}
    \log \Sigma = \log A + \beta \log D.
\end{equation}
This relation must be used with caution, since observational selection effects and biases are present in calibration samples (see Uro{\v s}evi{\' c} ~2002, ~2003, Uro{\v s}evi{\' c} et al.~2005, for more details). Another issue of this approach comes from the selection of the fitting method when calibrating the relation. For example, using vertical regression which minimizes deviations in $\Sigma$, will give result in a different correlation compared to the one which minimizes deviations in $D$. This relation was improved by using orthogonal regression, which has a symmetrical treatment of both variables, in Paper I and Pavlovi\'c et al.~(2014, hereafter Paper II). Also, Vukoti\'c et al.~(2014) showed that probability density function (PDF) of the calibration data in the fitting plain can be used to estimate distance-related properties of SNRs. Due to large scatter in the calibration data, as well as observational and fitting issues, it is important to update the calibration sample of SNRs with better constrained distances, whenever new observations are available. This will lead to more precise parameters which describe $\Sigma$ -- $D$ relation, which is often the only mean of determining the distance to many Galactic SNRs.

Following Paper I and Paper II, we present the updated empirical radio $\Sigma$ -- $D$ relation for Galactic SNRs. We extend our previous calibration sample, which now consists of $101$ shell-type SNRs and $9$ composite SNRs, where we can separate the flux from the surrounding shell from the flux in the central regions. We added $46$ new SNRs to our calibration sample and corrected distance estimates for $35$ SNRs (compared to Paper II). The fitting was done using orthogonal fitting, as well as cross validation kernel density smoothing (Duin 1976).

\section{Analysis and results}

\subsection{Calibration sample}

In order to derive new $\Sigma$ -- $D$ relation we use $110$ Galactic SNRs with available direct distance estimates. We use  $\Sigma$ at $1\ \mathrm{GHz}$, as was also done in  Papers I and II. Direct measurement of the surface brightness at this frequency is usually not available, so it is calculated based on measurements on other frequencies, and assuming a power law spectrum. Surface brightness values at $1\ \mathrm{GHz}$ were mostly taken from Green (
2019) catalog. List of all SNRs used in our calibration sample is given in Table~\ref{table:calibration}. Our new calibration sample has $46$ SNRs more than the Paper II sample, since more independent measurements of distances to SNRs are available. Some of the SNRs that were also in the calibration sample in Paper II now have a revised distance, based on recent, more precise measurements. All distances that we use, with their respective references, are given in Table~\ref{table:calibration}. Distances to SNRs in the calibration sample were determined from coincidences with the observed \hbox{H\,{\sc ii}} regions and molecular clouds, association with the red clump stars, pulsars, \hbox{H\,{\sc i}} absorption features and polarization,  H$\alpha$ line radial velocity or optical proper motion measurements.

\subsection{Composite SNRs}

{Although $\Sigma$ -- $D$ relation is predominantly used for SNRs with shell structure, some of SNRs which are characterized as composite can also contribute to the calibration of $\Sigma$ -- $D$ relation. Standard composite SNRs have nonthermal synchrotron centrally-filled character in the X-rays. Composite SNRs whose radiation from pulsars and/or pulsar-wind nebulae (PWNe) is either negligible in radio-wavelengths, or can be separated from shell radiation because of large spatial or spectral separation, can be used in our sample. In addition to these, plerionic composites, a growing class of so called mixed-morphology or thermal (X-ray) composite SNRs also exists (Rho and Petre 1998). This class of SNRs is characterized by the radio continuum shell-like structure with centrally brightened optically thin thermal X-ray emission. Mixed-morphology SNRs comprise a significant fraction of all the observed Galactic remnants and there are $33$ such objects in our calibration sample (see Table 1). They are known to expand in a very complex and usually high density environments (Zhang et al.~2015). For a significant number of these objects there is observational evidence of interaction with the adjacent \hbox{H\,{\sc i}} and molecular clouds (physical association with OH masers, coincidence with the CO distribution, strong infrared line emission etc.). Furthermore, the morphology of these remnants is often not in accordance with the standard evolutionary models that are commonly used to model shell SNRs (Zhang et al.~2019). The temperature across mixed-morphology is usually nearly uniform, and density, as well as pressure are mainly constant or sometimes increase toward the remnant center, which is not predicted by the commonly used Sedov-Taylor model. Lastly, one should bear in mind that some of the known mixed-morphology SNRs also host pulsar wind nebulae and/or detected pulsars, whose radiation can be separated from shell.

``Classical'' composite SNRs in our sample are $\mathrm{G}11.2-0.3$, $\mathrm{G}12.8$-$0.0$, $\mathrm{G}15.4+0.1$, $\mathrm{G}326.3-1.8$ and $\mathrm{G}338.3-0.0$, while thermal composite SNRs are $\mathrm{G}34.7-0.4$ (W44), 
G292.2$-$0.5, G8.7$-$0.1 and G93.3$+$6.9. Below, we shortly justify inclusion of mentioned SNRs in our calibration sample.

Borkowski et al.~(2016) refer to G11.2$-$0.3 as shell SNR, and note that its PWN dominates the X-ray image, and that it is hardly detectable in the radio image, with upper limit of $0.1$ mJy on the pulse flux obtained from deep $1.9$ GHz radio searches. G12.8$-$0.0 (W33) is a composite SNR that also has an X-ray pulsar, and  shell morphology in the radio-domain. G15.4$+$0.1 has X-ray PWN, coincident with TeV extended source, but has shell morphology in radio continuum. G326.3$-$1.8 (MSH 15-56, Kes 25) is composite SNR that contains a shell with a relatively steep radio spectrum and an interior flat-spectrum plerion (without any detected central pulsar). The flux density of the shell and plerion components can be separated (Dickel et al.~2000), so the shell can be used for calibration. Finally, G338.3$-$0.0 hosts an X-ray PWN, which has not been detected in the radio-domain. Data given for G338.3$-$0.0 in our Table 1 are from Castelletti et al.~(2011). They removed thermal component of the radiation of an SNR and put the upper limit to the flux density of any pulsar in the field of SNR to be $2$ and $1$ mJy for $610$ and $1280$ MHz, respectively.  Castelletti et al.~(2007) showed that  radio flux of PWN in W44 is lower than 0.1\% of  shell emission. Young pulsar J1119$-$6127 is positioned at the center of the SNR G292.2$-$0.5, with no significant radio emission detected from any PWN. The flux density at $1.374$ GHz for the pulsar is around $0.9$ mJy (Camilo et al.~2000), while the flux density of G292.2$-$0.5 at $1$ GHz is around $7$ Jy. Association of SNR G8.7$-$0.1 with the pulsar PSR B1800$-$21, inside western edge, is not clear (Brisken et al.~2006). Nevertheless, it has a flux density of order of mJy at $1.4$ GHz, while SNR G8.7$-$0.1 has a flux density around $80$ Jy at $1$ GHz (Ro\.{z}ko et al.~2018). G93.3$+$6.9 has well-defined shell-like morphology in the radio domain, and only after detection of an X-ray PWN (Jiang et al.~2007), it was suggested that this SNR hosts extended radio source, coincident with PWN. This central radio source is  detected in the $1.4$ GHz NRAO VLA Sky Survey (Condon et al.~1998) and has much lower flux density ($15$ mJy at $1.4$ GHz) compared to the flux density of the entire SNR ($9$ Jy).}

Finally, compared to our calibration sample from Paper II, we have excluded G189.1$+$3.0 (IC443), mixed-morphology SNR (that also has plerionic component), since its radio structure consists of two shells of different diameters, which are centered at different locations (Castelletti et al.~2011, Leahy~2004).

\vspace*{-0.4cm} 

\begin{center}
{\small
\setlength{\tabcolsep}{1pt}
\begin{longtable}{|c|l c c c c c|}
\caption{Calibration sample for $\Sigma$ -- $D$ relation, consisting of $110$ SNRs, with known distances.}\label{table:calibration}\\
\hline
\multicolumn{1}{|c|}{\textbf{No.}} & \multicolumn{1}{c}{\textbf{Catalog name}} &
\multicolumn{1}{c}{\textbf{Other name}} &\multicolumn{1}{c}{\textbf{Surface brightness}}  & \multicolumn{1}{c}{\textbf{Distance}} & \multicolumn{1}{c}{\textbf{Diameter}} &  \multicolumn{1}{c|}{\textbf{Dist.}}\\

\multicolumn{1}{|c|}{\textbf{}} & \multicolumn{1}{c}{\textbf{}} &
\multicolumn{1}{c}{\textbf{}} &
\multicolumn{1}{c}{\textbf{$(\times 10^{-21}\,\mathrm{W}\mathrm{m}^{-2}\mathrm{Hz}^{-1}\mathrm{sr}^{-1})$ }}  &
\multicolumn{1}{c}{\textbf{$\mathrm{(kpc)}$}} & \multicolumn{1}{c}{\textbf{$\mathrm{(pc)}$}} &  \multicolumn{1}{c|}{\textbf{Ref.}}\\ 
\hline 
\endfirsthead

\multicolumn{7}{c}%
{{\tablename\ \thetable{} -- Continued}} \\
\hline  \multicolumn{1}{|c|}{\textbf{No.}} & \multicolumn{1}{c}{\textbf{Catalog name}} &
\multicolumn{1}{c}{\textbf{Other name}} &
\multicolumn{1}{c}{\textbf{Surface brightness}}  & \multicolumn{1}{c}{\textbf{Distance}} & \multicolumn{1}{c}{\textbf{Diameter}} &  \multicolumn{1}{c|}{\textbf{Dist.}}\\

\multicolumn{1}{|c|}{\textbf{}} & 
\multicolumn{1}{c}{\textbf{}} &
\multicolumn{1}{c}{\textbf{}} &
\multicolumn{1}{c}{\textbf{$(\times 10^{-21}\,\mathrm{W}\mathrm{m}^{-2}\mathrm{Hz}^{-1}\mathrm{sr}^{-1})$ }}  & \multicolumn{1}{c}{\textbf{$\mathrm{(kpc)}$}} & \multicolumn{1}{c}{\textbf{$\mathrm{(pc)}$}} &  \multicolumn{1}{c|}{\textbf{Ref.}}\\ 

\hline 
\endhead

\multicolumn{7}{c}{\parbox{\textwidth}{\small \textbf{Notes:} $^{a}$ SNRs with the revised distances, compared to Paper II; $^{b}$ New SNRs, not used in the calibration samples in Papers I and II; $^{c}$ SNRs removed due to poor quality distances; $^{d}$ SNRs removed due to possibly being in the free expansion phase; $^{e}$ Mixed-morphology SNRs (Ferrand and Safi-Harb~2012, Zhang et al.~2015).\\
\textbf{References}: (1) Sankrit et al.~2016; (2) Vel\'azquez et al.~2002; (3) Hewitt and Yusef-Zadeh 
 2009; (4) Shan et al.~2018; (5) Minter et al.~2008; (6) Halpern et al.~2012; (7) Su et al.~2017a; (8) Ranasinghe and Leahy 2018a; (9) Ranasinghe and Leahy 2018b; (10) Ranasinghe et al.~2018; (11) Su et al.~2014; (12) Yamaguchi et al.~2004; (13) Giacani et al.~1998; (14) Ranasinghe and Leahy 2017; (15) Matthews et al.~1998; (16) Tian and Leahy 2006; (17) Lozinskaya et al.~1993; (18) Fesen et al.~2018; (19) Leahy and Green 2012; (20) Jackson et al.~2008; (21) Foster and Routledge 2003; (22) Uyanıker et al.~2002; (23) Forster et al.~2004; (24) Kothes et al.~2005; (25) Tian et al.~2007; (26) S\'anchez-Cruces et al.~2018; (27) Alarie et al.~2014; (28) Yar-Uyanıker et al.~2004; (29) Pineault et al.~1993; (30) Zhang et al.~2013; (31) Leahy and Tian 2006; (32) Zhou et al.~2016; (33) Foster et al.~2013; (34) Katsuda et al.~2016; (35) Leahy and Tian 2007; (36) Arias et al.~2019; (37) Chatterjee et al.~2009; (38) Su et al.~2017b; (39) Yu et al.~2019; (40) Reynoso et al.~2017; (41) Katsuda et al.~2008; (42) Kamitsukasa et al.~2016; (43) Shan et al.~2019; (44) Reynoso et al.~2006; (45) Caswell et al.~2004; (46) Giacani et al.~2000; (47) Prinz et al.~2013; (48) Gaensler et al.~1998; (49) Sawada et al.~2019; (50) Andersen et al.~2011; (51) Doherty et al.~2003; (52) Caswell et al.~1975; (53) Rosado et al.~1996; (54) McClure-Griffiths et al.~2001; (55) Nikoli\'c et al.~2013; (56) Vink 2004; (57) Zhu et al.~2015; (58) Eger et al.~2011; (59) Frail et al.~1996; (60) Takata et al.~2016; (61) Supan et al.~2018; (62) Lemiere et al.~2009; (63) Kothes and Dougherty 2007; (64) Giacani et al.~2011; (65) Yamauchi et al.~2013; (66) Moriguchi et al.~2005; (67) Tian and Leahy 2012; (68) Tian and Leahy 2014; (69) Tian et al.~2007; (70) Giacani et al.~2009; (71) Tian et al.~2010; (72) Bamba et al.~2000.}}\\
\endlastfoot

\hline
 1 & G4.5$+$6.8$^{a,d}$     &  Kepler, SN1604, 3C358  &   317.8   &   5.1  &   4.4   &   1   \\
 2 & G6.4$-$0.1$^{b,e}$     &  W28  &   310   &  1.9  &   26.5   &  2   \\
 3 & G8.7$-$0.1$^{b,e}$     & W30  &    80  &  4.5  & 58.9    &  3   \\
 4 & G9.7$-$0.0$^{b}$      &    & 3.3    &  4.7   &  17.6    &   3   \\
 5 & G11.0$-$0.0$^{b}$   &    &    2.0   & 2.4     &   6.9  &   4   \\
 6 & G11.2$-$0.3$^{a,c}$     &    &  22    &  6.25  &  7.3    &  5   \\
 7 & G12.8$-$0.0$^{b,c}$     & W33   & 0.8     & 4.8   &  4.2    &  6   \\
 8 & G15.1$-$1.6$^{b,c}$     &    &     1.1  & 2.2     &    17.2  &    4  \\
 9 & G15.4$+$0.1$^{b}$     &      &    5.6    &    9.3     &   39.2      &     7  \\
 10 & G18.1$-$0.1$^{a}$    &    &     10.8  & 6.4     &     14.9 &  8    \\
 11 & G18.6$-$0.2$^{b}$     &    &   5.8   & 4.4    &     7.7   &    8  \\
 12 & G18.8$+$0.3$^{a}$     &  Kes 67  &  25.3    &    13.8 &  54.9   &    8  \\
 13 & G21.8$-$0.6$^{a,e}$     &  Kes 69  &  24.5    &  5.6  &  32.6    & 8     \\
 14 & G22.7$-$0.2$^{b}$     &    &    7.3  &  4.7   &    35.5  &     8 \\
 15 & G23.3$-$0.3$^{a}$     &  W41  &    14.4  &  4.8   &     37.7 &  8    \\
 16 & G24.7$-$0.6$^{b}$     &    &    5.4  & 3.8    &    16.6  &  9     \\
 17 & G27.4$+$0.0$^{a}$     & Kes 73, 4C--04.71   &    56.4  & 5.8 &  6.7   &   8  \\
 18 & G28.6$-$0.1$^{b}$     &    &    3.7  &  9.6   &   30.2   &  9    \\
 19 & G29.6$+$0.1$^{b}$     &    &    9.0  &   4.7  &    6.8  &  10    \\ 
 20 & G31.9$+$0.0$^{e}$     &  3C391  &  104.5    &  7.2   &  12.4    &   11   \\
 21 & G32.4$+$0.1$^{b,c}$     &    &    1.0  & 17   & 29.7     &  12   \\
 22 & G32.8$-$0.1$^{b}$     &  Kes 78  &    5.7  &   4.8  &    23.7  & 8     \\ 
 23 & G33.6$+$0.1$^{a,e}$     &  Kes 79, 4C00.70, HC13  & 30.1     &  3.5   &  10.2    &   8   \\
 24 & G34.7$-$0.4$^{b,e}$   &  W44   &  38.2    &   3.0  &   26.8   &   8   \\  
 25 & G35.6$-$0.4$^{a}$     &    &    8.0  & 3.8    &  14.2      &   8   \\
 26 & G41.1$-$0.3$^{a,e}$     & 3C397   &  307.2    &    8.5  &  8.3    &    8  \\
 27 & G41.5$+$0.4$^{b}$     &    &    1.5  &   4.1  &    11.9  &   10   \\
 28 & G43.3$-$0.2$^{a,e}$     &  W49B  &    467.0  &  11.3   &    11.4  &  8    \\
 29 & G46.8$-$0.3$^{c}$      &  HC30  &    11.4  &  8.55   &    37.0  &   8   \\
 30 & G49.2$-$0.7$^{b,e}$    &  W51  & 26.8     &   5.4  &     47.1 &   8   \\
 31 & G53.6$-$2.2$^{e}$     & 3C400.2, NRAO 611   &  1.3    &  2.8   &  24.8    &   13   \\ 
 32 & G54.4$-$0.3$^{a}$     &  HC40  &    2.6  &  6.6   &    76.8  &   14   \\
 33 & G55.0$+$0.3     &    &    0.2  & 14    & 70.5     &   15   \\
 34 & G57.2$-$0.8$^{b,c}$     & 4C21.53   &    1.9  &   6.75  &   23.6   &   10   \\
 35 & G65.1$+$0.6$^{a}$     &    &    0.2  &   9.2  &    179.5  &   16   \\
 36 &   G67.7$+$1.8$^{b}$     &    &  0.8    & 2    &  7.8    &   4   \\
 37 &   G73.9$+$0.9$^{b,c}$     &    &     1.8 &  1.25   &  9.8    &   17   \\
 38 &   G74.0$-$8.5$^{a}$     &  Cygnus Loop  & 0.8     &  0.74   &  41.0    &  18    \\
 39  & G78.2$+$2.1$^{a}$  &  DR4, gamma Cygni SNR &   13.4   &   1.9 &   33.2   & 4    \\
 40 &   G82.2$+$5.3$^{b,e}$     & W63   &    2.8  &   3.2  &    73.1  &  4    \\
 41 &   G84.2$-$0.8     &    &    5.1  &  6   &  31.2    &   19   \\ 
 42 &   G85.4$+$0.7$^{b,e}$     &    &   0.6   &   4.4  &    30.7  &  4    \\\hline
 43 &   G85.9$-$0.6$^{b,e}$     &    &  0.6    & 4.8    &    33.5  &   20   \\
 44 &   G89.0$+$4.7$^{a,e}$     &  HB21  &  3.0    & 1.9    &    57.4  & 4     \\
 45 &   G93.3$+$6.9$^{e}$   &   DA 530, 4C(T)55.38.1  &   2.4    &    2.2   & 14.9    &  21  \\
 46 &   G93.7$-$0.2$^{e}$     & CTB 104A, DA 551   &  1.5    &   1.5  &    34.9  &   22   \\
 47 &   G94.0$+$1.0$^{a}$     & 3C434.1   &     2.6 &   5.2  & 41.4     & 23     \\
 48 &   G96.0$+$2.0     &    &    0.1  & 4    & 30.2     &   24   \\
 49 &   G108.2$-$0.6     &    &    0.3  &  3.2   &  57.2    &  25    \\
 50 &   G109.1$-$1.0$^{a,e}$     &  CTB 109  &    4.2  & 3.1    &    25.2  &      26 \\
 51 &   G111.7$-$2.1$^{d}$    &  Cassiopeia A, 3C461  &   11452.7   & 3.33    &    4.8  &    27  \\
 52 &   G114.3$+$0.3     &    &  0.2    & 0.7    &     14.3 &   28   \\
 53 &   G116.5$+$1.1    &    &    0.3  &  1.6   &    32.2  &   28   \\
 54 &   G116.9$+$0.2$^{e}$  &  CTB 1  &  1.0    &  1.6   &   15.8   & 28     \\
 55   &  G119.5$+$10.2   &   CTA 1      &   0.7   &     1.4  &    36.6   &   29   \\
 56 &   G120.1$+$1.4$^{d}$     & Tycho, 3C10, SN1572   &    131.7  & 2.5    &  5.8   & 30 \\
 57 &   G127.1$+$0.5$^{a}$     &  R5  & 0.9     &    1.15 &    15.0  &  31    \\
 58 &   G132.7$+$1.3$^{a,e}$     &  HB3  &   1.0   &  1.95   &    45.4  &  32    \\
 59 &   G152.4$-$2.1$^{a}$     &    &    0.06  &  1.1   &    31.2  &   33   \\
 60 &   G156.2$+$5.7$^{a,c,e}$     &    &    0.1  &    1.7 &    54.4  &    34  \\
 61 &   G160.9$+$2.6$^{e}$     &  HB9  &    1.0  &  0.8   &    30.2  &   35   \\
 62 &   G166.0$+$4.3$^{a,e}$     &  VRO 42.05.01  &  0.5    &    1.0 &  12.8    &  36    \\
 63 &   G180.0$-$1.7$^{a}$    &    S147     &      0.3       &    1.33     &   69.6       &   37     \\
 64 &   G190.9$-$2.2     &    &   0.05   &   1  &    18.8 &   33   \\
 65 &   G205.5$+$0.5$^{a,c}$     &  Monoceros Nebula  &    0.4  &  1.6   &   102.4   & 38      \\
 66 &   G206.9$+$2.3$^{b,c}$     & PKS 0646+06   &  0.4    &    1.6 &    22.8  &     38 \\
 67 &   G213.0$-$0.6$^{b}$     &    &    0.14  &  1.15   &    50.1 &  39   \\
 68 &   G260.4$-$3.4$^{a}$     & Puppis A, MSH 08-44   &    6.5  &    2.2 &    35.1  &  40    \\
 69 &   G266.2$-$1.2$^{b}$     &  Vela Jr, RX J0852.0--4622  &    0.5  & 0.75    &  26.2    &  41    \\
 70 &   G272.2$-$3.2$^{b,c,e}$     &    &    0.3  &  2.5   &   10.9   &   42   \\
 71 &   G279.0$+$1.1$^{b}$     &    &    0.5  &    2.7 &    74.6  &   43   \\
 72 &   G284.3$-$1.8$^{b}$     & MSH 10-53   &    2.9  & 5.5     &  38.4    &   43   \\
 73 &   G290.1$-$0.8$^{e}$     &  MSH 11-61A  & 23.2     &  7   &  33.2    &   44   \\ 
 74 &  G292.2$-$0.5$^{e}$ &                  &      3.5  &        8.4      &  42.3     &  45   \\
 75 &   G296.1$-$0.5$^{b}$     &    &    1.2  &  4.3   &   38.0   &    43  \\
 76 &   G296.5$+$10.0    &  PKS 1209-51/52  & 1.2     & 2.1    & 46.7     &    46  \\
 77 &   G296.7$-$0.9    &    &  3.8    &    9.8 &     31.2 & 47     \\
 78 &   G296.8$-$0.3     & 1156-62   &    4.7  & 9.6    &    46.7  &   48   \\
 79 &   G299.2$-$2.9$^{b}$     &    &   0.4   &    2.8 &    11.5  &   43   \\
 80 &   G306.3$-$0.9$^{b,c}$     &    &    1.5  &   20  &    23.3  &    49  \\
 81 &   G308.4$-$1.4$^{a}$     &    &    0.7  &    3.1 &    7.6  &   43   \\
 82 &   G309.2$-$0.6$^{b}$     &    &     5.8 &    2.8 &    10.9  &  43    \\
 83 &  G311.5$-$0.3$^{b,e}$      &    &     18.1    &   14.8      &   21.5     &     50   \\
 84 &   G312.4$-$0.4$^{b,c}$     &    &     4.7 &   14  &    154.8  &    51  \\
 85 &   G315.4$-$2.3$^{a,c}$    &  RCW 86, MSH 14-63  &  4.2    &   2  &     24.4 &   43   \\
 86 &   G316.3$-$0.0$^{b,c}$     &  MSH 14-57  &  6.5    &    7.2 &   42.2   &   52   \\
 87   & G326.3$-$1.8$^{b}$   &  MSH 15-56, Kes 25  &   11.9   &   4.1  &   45.3   &    53    \\
 88 &   G327.4$+$0.4$^{a,e}$     &  Kes 27  &     10.2 & 4.3    & 26.3     &  54    \\
 89 &   G327.6$+$14.6$^{a,c}$    &  SN1006, PKS 1459-41  &    3.2  & 2    &  17.4    &  55    \\
 90 &   G330.2$+$1.0$^{b}$     &    &    6.2  &  4.9   &    15.7  &   54   \\
 91 &   G332.4$-$0.4$^{a}$     & RCW 103   &    42.1  & 3    &   8.7   &   43   \\
 92 &   G332.4$+$0.1$^{b,c}$     &  MSH 16-51, Kes 32  &    17.4  &  7.5   &    32.7  &  56    \\
 93 &   G332.5$-$5.6$^{b}$     &    &    0.2  &    3.0 &   30.5   &   57   \\
 94 &   G335.2$+$0.1$^{b}$     &    &    5.5  & 1.8    &    11.0  &  58  \\
 95 &   G337.0$-$0.1     &  CTB 33  & 100.4     & 11    &   4.8   &    59  \\
 96 &   G337.2$-$0.7$^{b,c}$    &    &    6.3  & 9   &    15.7  &  60    \\
 97 &   G337.8$-$0.1$^{a,e}$     &  Kes 41  &   48.2   &  12   &  25.6    & 61     \\
 98  & G338.3$-$0.0$^{a}$     &    &    16.5  &  10.5  &  24.4  &  62   \\
 99 &   G340.6$+$0.3     &    &    20.9  &  15   &    26.2  &  63    \\
 100 &   G344.7$-$0.1$^{e}$     &    &   5.9   &  6.3   &  14.7    &   64   \\
 101 &   G346.6$-$0.2$^{e}$     &    &    18.8  &    7.5 &    17.4 &   65   \\
 102 &   G347.3$-$0.5$^{b}$     &  RX J1713.7-3946  &    1.2  &    1 & 17.4     &   66   \\
 103 &   G348.5$+$0.0$^{b,c}$     &    &    15.0  &    6.3 &   18.3   &   67   \\
 104 &   G348.5$+$0.1$^{a,e}$     &  CTB 37A   & 48.2     &    7.9 &  34.5    &    67  \\
 105 &   G348.7$+$0.3     &  CTB 37B  &    13.5  &   13.2  &    65.3  &     67 \\
 106 &   G349.7$+$0.2    &    &    594.8  &  11.5   &   7.5   &  68    \\
 107 &   G351.7$+$0.8$^{b}$     &    &    5.9  &  13.2   &  61.0    &  69    \\
 108 &   G352.7$-$0.1$^{e}$    &    &    12.3  &  7.5   &    15.1  & 70     \\
 109 &   G353.6$-$0.7$^{b}$     &    &    0.4  &    3.2 &    27.9  &  71    \\
 110 & G359.1$-$0.5$^{a,e}$   &    &   3.6      &    8.5     &    59.3     & 72       \\
\hline
\end{longtable}
}
\end{center}

\newpage

\centerline{
\includegraphics[width=0.23\columnwidth]{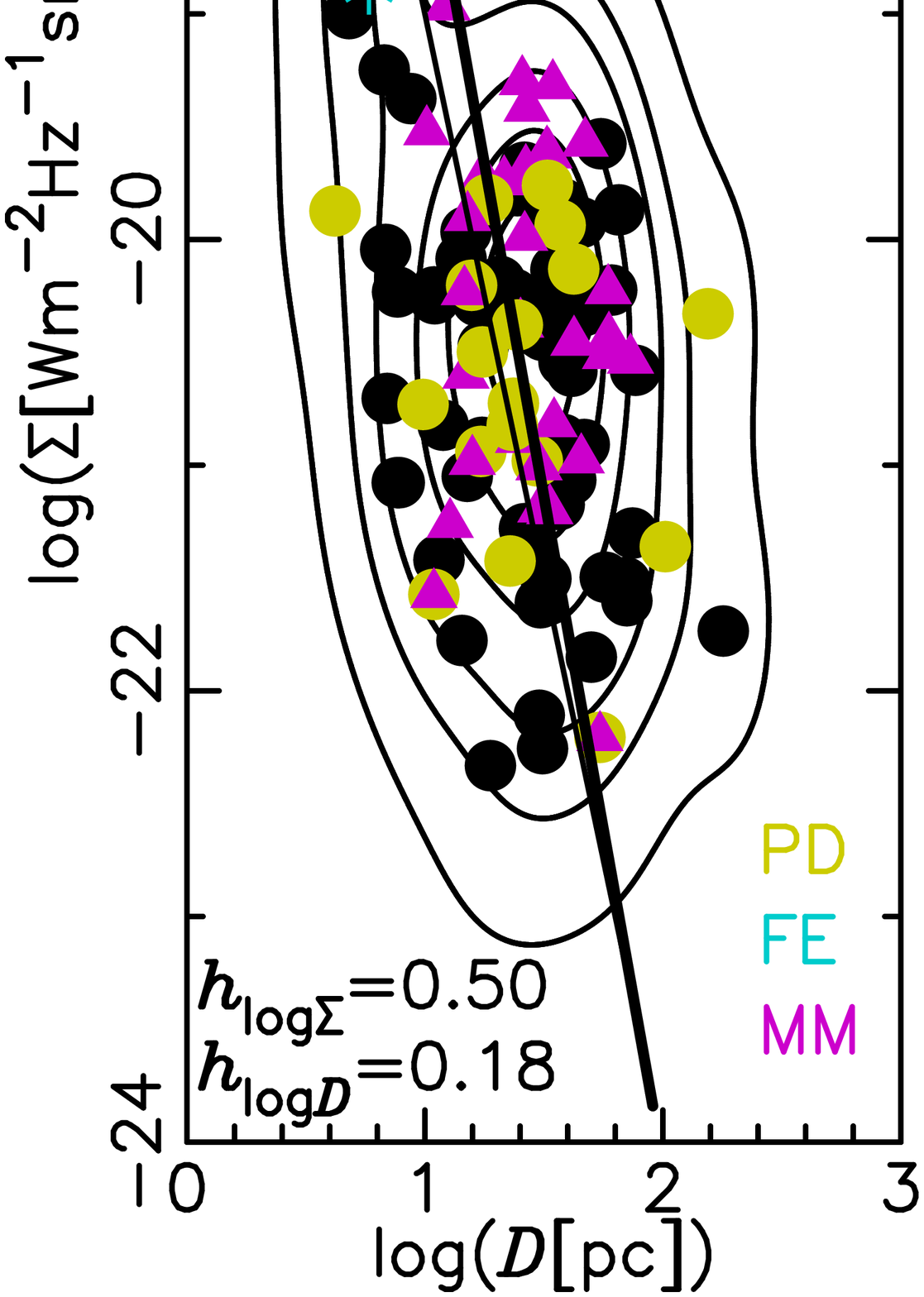}
\hspace{0.5cm}
\includegraphics[width=0.23\columnwidth]{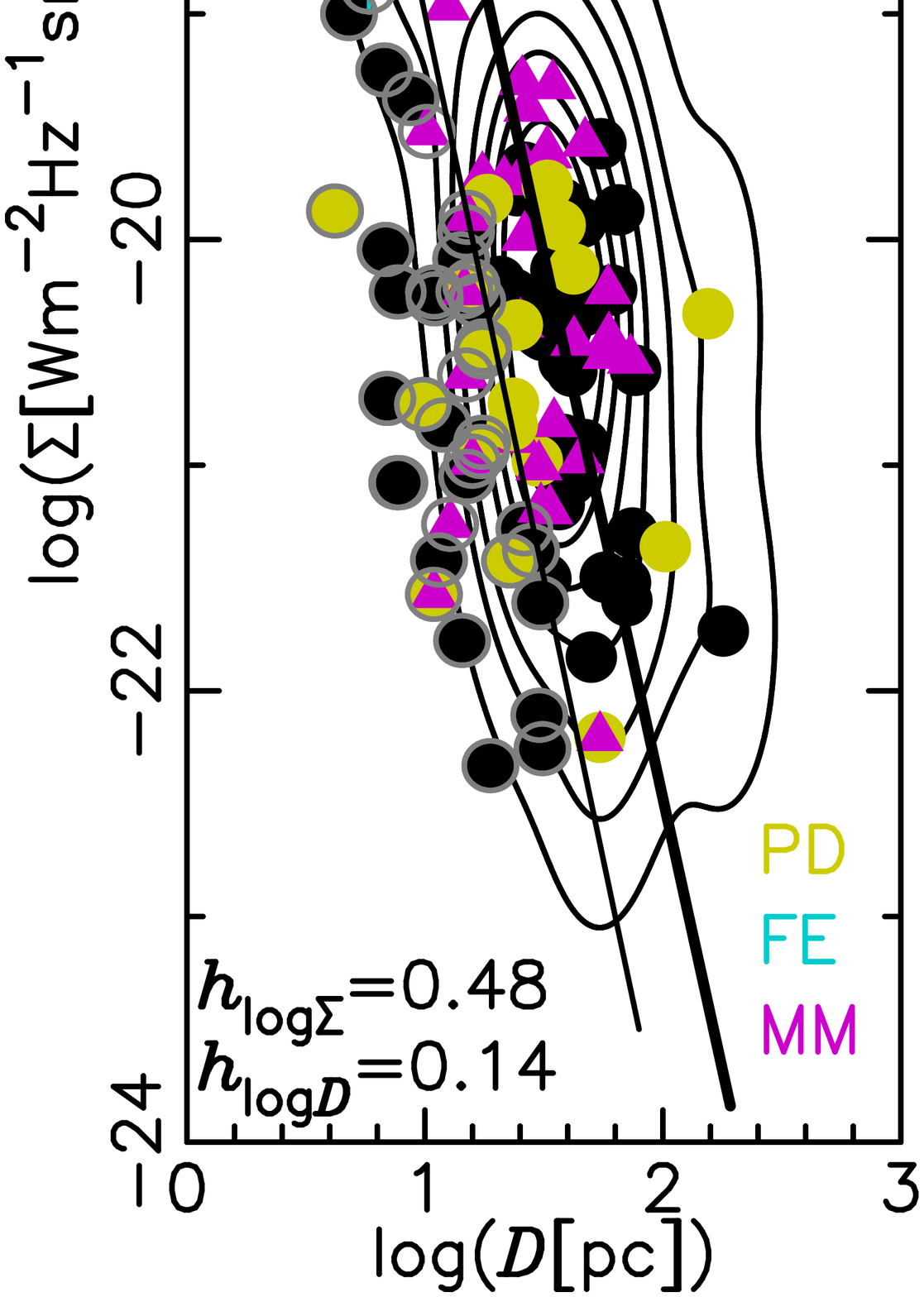}
\hspace{0.5cm}
\includegraphics[width=0.23\columnwidth]{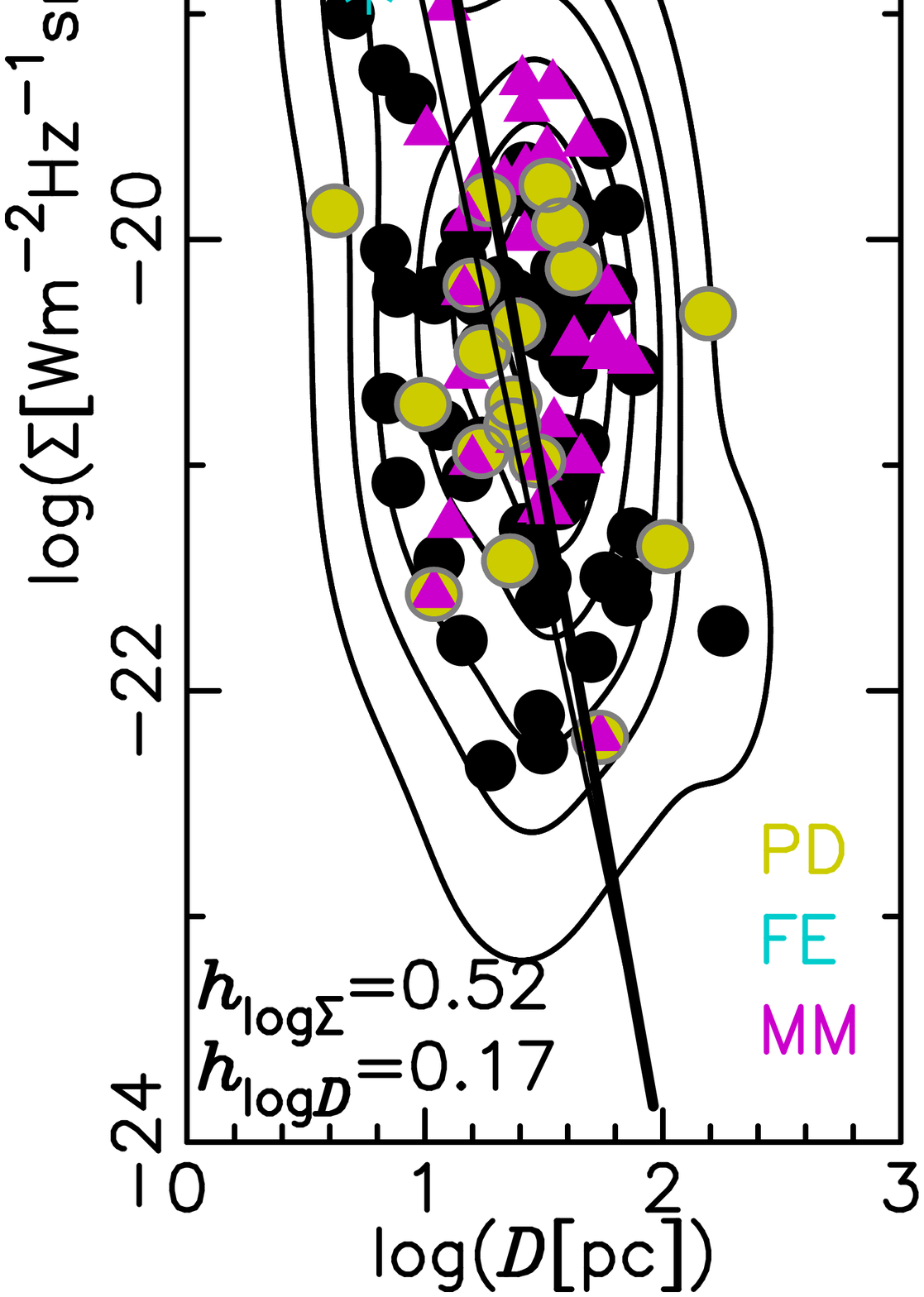}
}
\vskip2cm
\centerline{
\includegraphics[width=0.23\columnwidth]{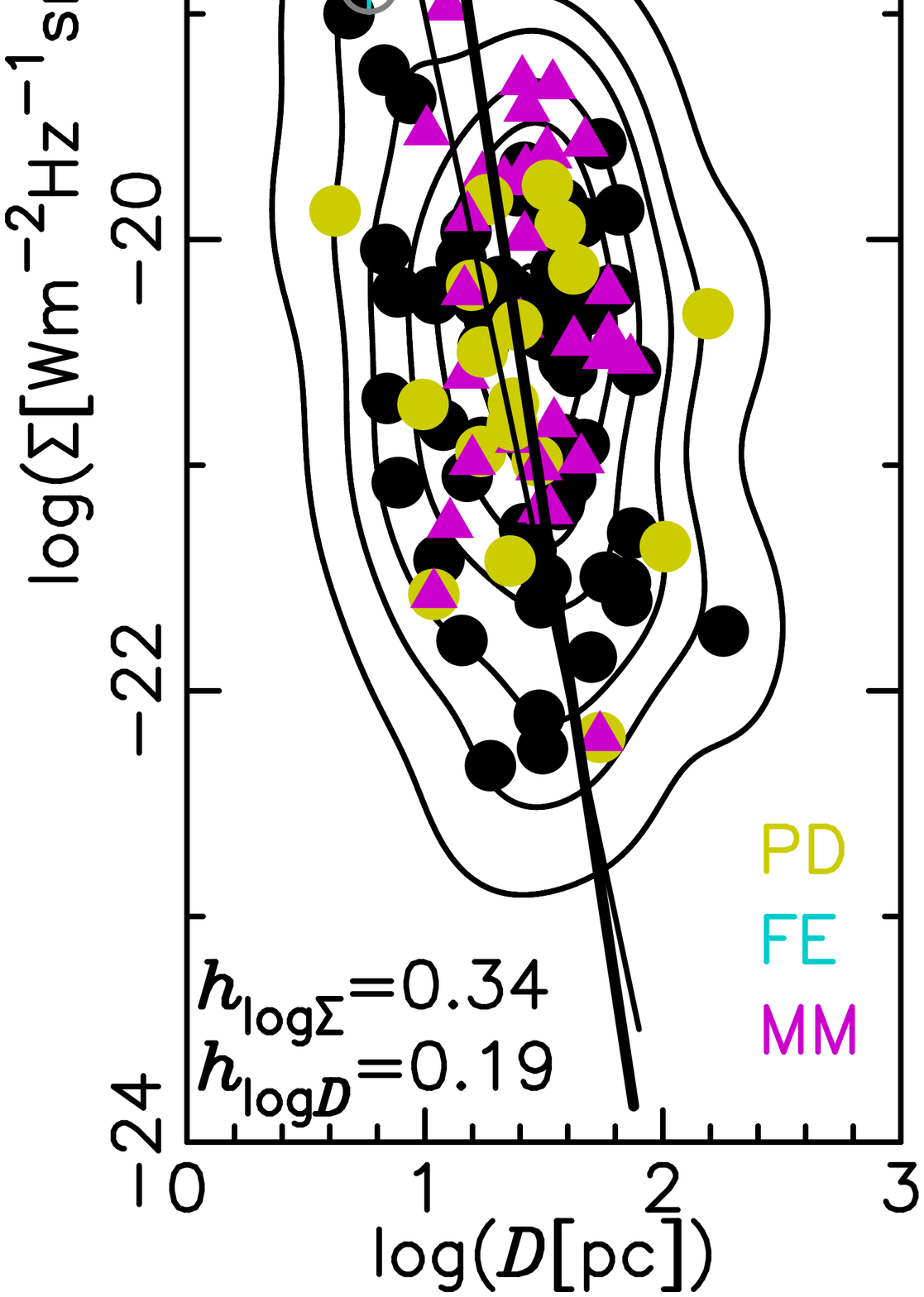}
\hspace{0.5cm}
\includegraphics[width=0.23\columnwidth]{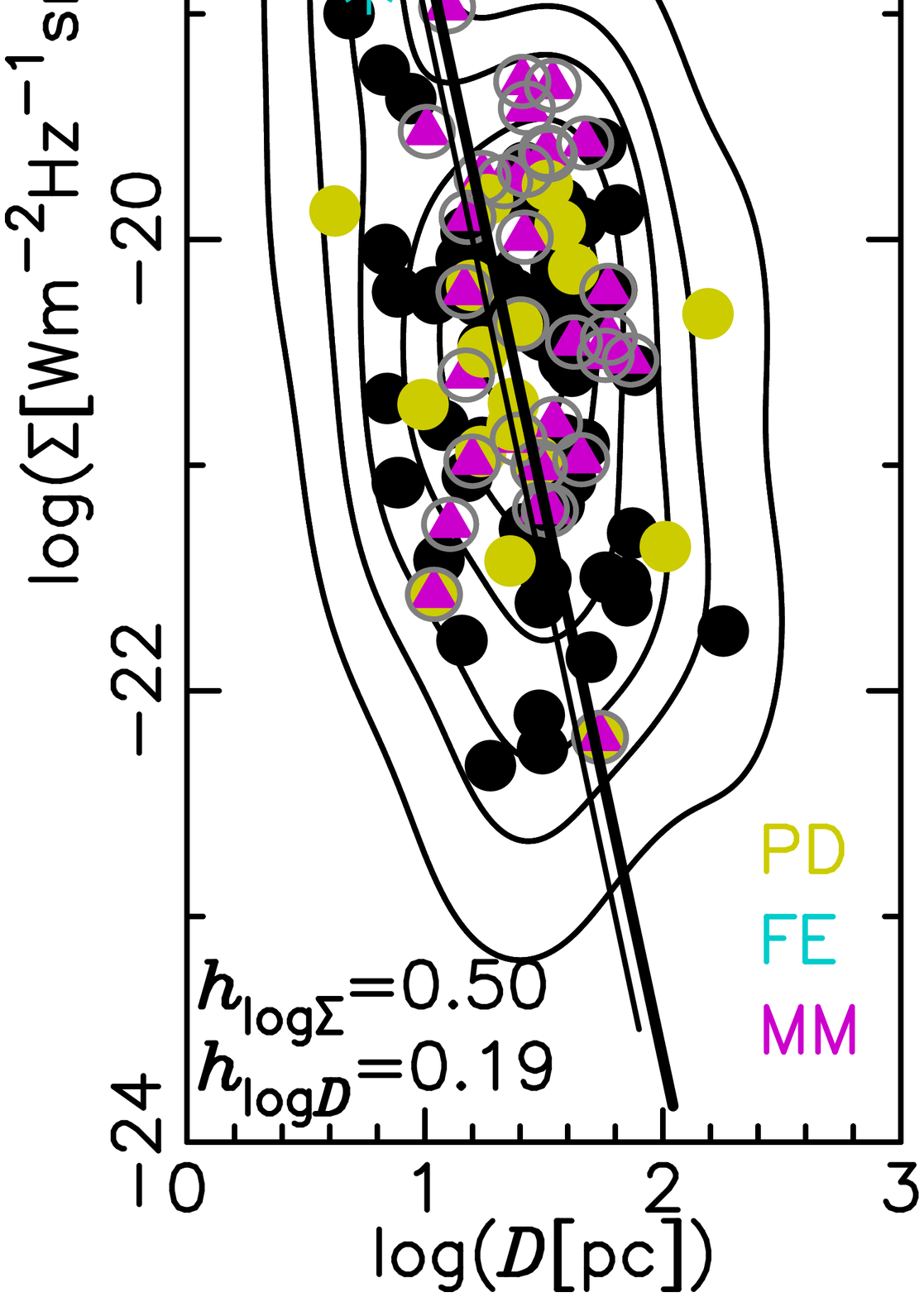}
\hspace{0.5cm}
\includegraphics[width=0.23\columnwidth]{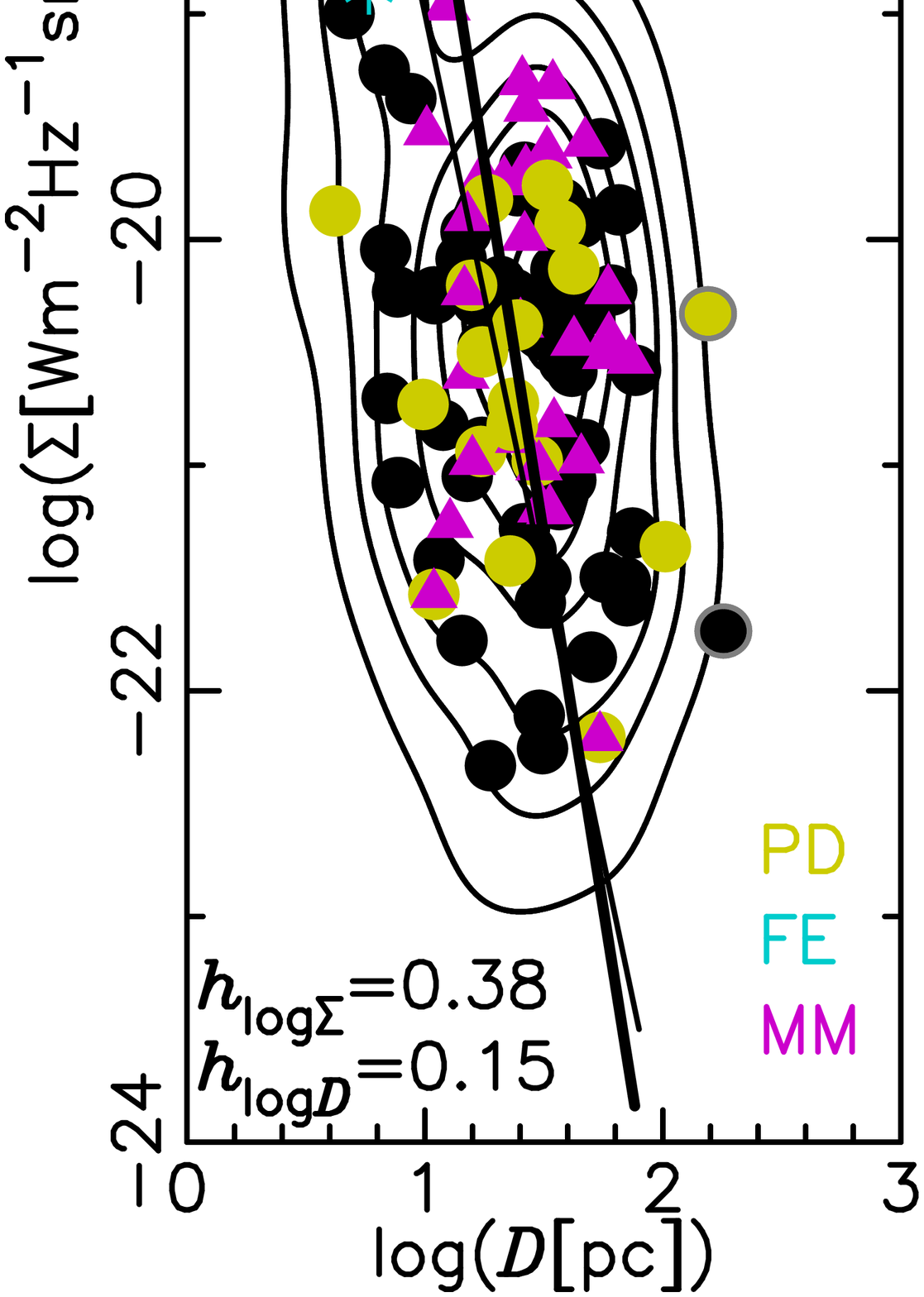}
}
{\bf Figure 1.}{\footnotesize The updated sample of Galactic SNRs and applied filters. Filter names are given in the title of each panel. The maximum likelihood kernel density data smoothing in $\log \Sigma-\log D$ with designated bandwidths $h$ is presented with contour levels at $0.01,~ 0.05,~ 0.1,~ 0.2,~ 0.3,~ 0.4$. The corresponding orthogonal fit (thick solid line) parameters and their resulting distance fractional errors ($\mathbf{f}$) are also designated on each panel. The slope of the fit line is also given in degrees ($\beta[^o]$, angle between the fit line and $-\log D$ axis). Data points: PD -- Data points with poor distances (yellow circles), FE -- objects in free expansion phase (blue asterisks), MM -- mixed-morphology remnants (violet triangles), or black circles otherwise. Data points that are filtered out from the sample are marked with open gray circles. The number of filtered out data points is presented as a gray color number next to the total number of the points in the non-filtered sample (in the top right corner of each panel). The cut-off line for the simulation based filter (see Section 2.5), according to the results of Pavlovi\'c et al.~(2018), is shown as a thin solid line with $\beta = -5$, which contains the point $\log \Sigma = -19 $ and $ \log D  = 1.0 $. For details on kernel smoothing filter see sections 2.4 and 2.5.}

\newpage

\centerline{
\includegraphics[width=0.75\columnwidth]{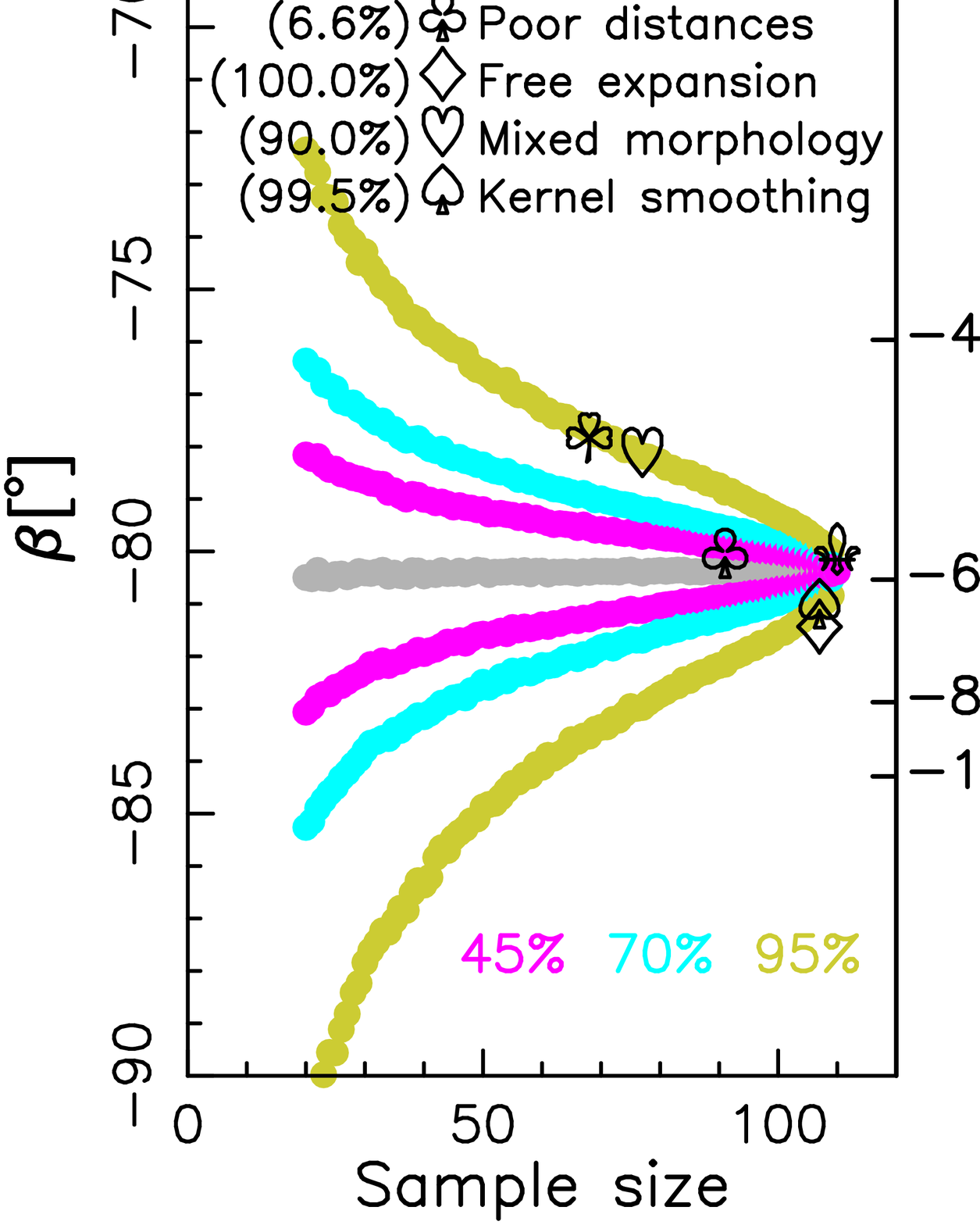}
}
{\bf Figure 2.}{\footnotesize Orthogonal fit slope confidence map for the selected sample of $110$ Galactic SNRs. The sample is randomly re-sampled without repetition of individual points. For size of the re-samples, designated on the horizontal axis, the orthogonal fit slope for the $10^4$ re-samples was calculated. The median slope value of these slopes is plotted with gray dots, while the confidence intervals around the median are color-coded as designated -- the percentages indicate the ratio of re-samples that falls  in the interval around the median specified with the plotted dots. In the same manner, each filtered sample is presented as designated on the plot, at the given value of the  confidence interval (in brackets).}

\subsection{Orthogonal fitting}

For the purpose of $\log \Sigma-\log D$ slope calculation we used fitting procedure with orthogonal offsets of the data points from the fit line. Statistical tests from Paper I (see also Keles 2018) confirmed the robust nature of orthogonal fitting when compared with fitting procedures that use different types of offsets and methods for linear fit slope calculation. Also, the orthogonal approach has superior performance over vertical offsets fitting in the case of extra-galactic M82 sample of SNRs (Uro\v sevi\'c et al.~2010).

The uncertainties of parameters for the orthogonal  $\log \Sigma = \log A + \beta * \log D $  fit are estimated from a simple bootstrap procedure (Efron and Tibshirani 1994). For each fitted sample, the $1000$ bootstrap re-samples with replacement are performed, using SFMT pseudo-random number generator (Saito and Matsumoto 2008). Each of the $1000$ sets of fit parameters are stored in an array and the uncertainties are estimated as standard deviation of that array. Based on the various selection criteria we apply this procedure to several sub-samples and the results are presented in Figs.~1 and 2 (see the following text, sections 2.4 and 2.5, for the detailed explanations of the figures).

\subsection{Kernel density smoothing}

The application of density smoothing in $\log \Sigma - \log D$ plane (Vukoti\'c et al.~2014) showed that such 2D data density distribution can reveal more information about the data sample than just the parameters of the best fit line, when only fitting is used. The procedure was further improved with bootstrap based kernel density smoothing (Bozzetto et al.~2017) and kernel bandwidth ($h$) selection using maximum likelihood cross-validation (1D procedure described in Maggi et al.~2019).

The later was modified to a 2D case with a simple product Gaussian kernel, for XY data usually written as $K(X,h_\mathrm{X})\cdot K(Y,h_\mathrm{Y})$. This type of kernel is applied to the updated Galactic SNR sample of $110$ objects. The contours in Figs.~1 and 2 are calculated for the optimal kernel bandwidths of $h_\mathrm{\log \Sigma}= 0.52$ and $h_\mathrm{\log D}= 0.175$. 

\subsection{Sample filtering}

From the upper left panel of Fig.~1 it is evident that the full sample of $110$ compiled objects has a slope of $-6$, which is steeper  when compared with the recent empirical and theoretical findings; Paper I and Paper II report empirical slope of  around $-5$, while Pavlovi\'c et al.~(2018) reported slopes between $-4$ and $-6$, depending on the density of the ISM in the simulated evolution trajectories. Such a steep slope implies a very low correlation of the $\Sigma D$ data and underlines the need for further scrutiny of the $\Sigma-D$ calibration samples. To obtain a shallower slope and  better agreement with present theoretical findings of SNR evolution, we test various filtering criteria on the selected samples of calibrators.

Pavlovi\'c et al.~(2018) simulated the evolution trajectories of SNRs applying different values for density of the environment, and initial explosion energy. Their Fig.~3 shows that the considered fiducial density values of the environment give the slope values in the $-4$ (low density) to $-6$ (high density) interval. In addition, they plot the sample of $65$ Galactic SNRs over the simulated evolution trajectories. Some of the objects with high values of $\Sigma$ are located above the evolution trajectories while even more of them, with the low $\Sigma$ values, are below the simulated trajectories. This indicates that the empirical slope might be steeper than the slopes obtained from simulations. The low brightness and small diameter part of the sample might be dominated by objects that evolve in high density environments and have low brightness which makes them susceptible to sensitivity related selection effects and can also contribute to overall scatter of the relation. Applying the simulation based filter described below, we remove these objects from the sample and discuss their influence on the agreement between the theoretical and empirical slopes.

We apply, and later discuss, several filters, along with the full sample, to analyse and trace the possible cause of discrepancy between empirical and simulated data  (Figs.~1, 2 and Table 2):\\
\indent(i) All data points,\\
\indent(ii) Simulation based filter,\\
\indent(iii) Poorly determined distances,\\
\indent (iv) Free expansion filter,\\
\indent(v) Mixed-morphology and\\
\indent(vi) Kernel smoothing outliers.\\
The orthogonal fitting and kernel density smoothing, as described above (sections 2.3 and 2.4), are applied to each filtered sample and the results are presented in Figs.~1 and 2 and summarized in Table 2.
{In Fig.~1 we present the updated sample of Galactic SNRs with applied filters, as designated in the title of each panel. The maximum likelihood kernel density data smoothing in $\log \Sigma - \log D$, with designated bandwidths
$h$, is presented with contours. The corresponding orthogonal fit (thick
solid line) parameters and their resulting distance fractional errors ($f$) are also given on each panel. The data points are designated as follows: PD -- data points with poor distances (yellow circles), FE -- objects in free expansion phase (blue asterisks), MM -- mixed-morphology remnants (violet triangles), or black circles otherwise. Data points that are filtered out from the sample are marked with open gray circles.}

The simulation filter is the most radical in terms of the number of expelled points from the sample and also gives the shallowest slope. The filter selects all data points that are above the line of $\beta = -5$ and contains the point $\log \Sigma = -19 $ and $ \log D  = 1.0$. This filters out data points that are positioned below the simulated evolution trajectories in Pavlovi\'c et al.~(2018) -- as mentioned above, these are mostly the low brightness and small diameter objects that are likely to be subjected to sensitivity related selection effects.

Conversely, the exclusion of objects with poorly determined distances {(objects with large or no distance errors, only upper or lower limits to distances available, or very different distances derived using different methods)}, more than $\approx$17\% of the total number of objects, seems to have no significant effect on any of the sample properties. Deselecting $3$ objects in the free expansion phase, or $3$ objects that are outside the second contour level (contour level of $0.05$ (Fig.~1, bottom row, left and right panel, respectively), gives steeper slope of $-7$. Finally, the exclusion of $33$ mixed-morphology remnants from the sample gives a shallower slope of $-5$, similar as the simulation based filter, because the MM SNRs should be mainly the low brightness and small diameter objects. 

However, the slopes from all filters correspond to the $\approx -80^\circ$ angle of the fit line to the horizontal axis (Table 2). Since the slope of the fit line is the tangent function of this angle, the  resulting changes of the fit line slopes do not significantly reflect the overall stretch in the direction of the sample. The tangent function of such an angle  tends to infinity as the angle approaches $\pm \pi/2$. For these reasons, better insight is achieved when the slope is expressed as the above mentioned angle. The bootstrap confidence estimates (Section 2.3) are performed on this angle variable because, unlike the slope, it is continuous over angle values of $\pm \pi/2$ and gives meaningful results. The results are then translated back to the slope variable. It is indicative that values for a clockwise angle ($\beta[^\circ]$) with the $-\log D$ axis overlap within the estimated uncertainties, for all filters. 

This means that no change of the slope, by any of the filters, is to be considered as statistically significant, but nevertheless, the difference in slope values and other parameters of the filters (e.g.~distance fractional error) can be used to discuss the consistency of the samples in order to make a more reliable $\Sigma-D$ calibration (Section 3).

\section{Discussion}

From Figs.~1 and 2 it is evident that the simulation  based filter (see point ii in Section 2.5) is the most effective in reducing the steep slope of the full sample. The {\sl a priori} slope of the filter line of $-5.0$ is the same as the slope of the filtered sample and both are in agreement with the theoretical insights of SNRs evolution in Sedov-Taylor phase. At first glance this agreement may seem obvious, and without added value to the present knowledge on the subject, but a careful inspection does point out to some possibilities that might improve the $\Sigma-D$ relation. It indicates that the ``upper'' part of the sample, the one above the simulated evolutionary lines from Pavlovi\'c et al.~(2018), does not affect the slope of the relation as much, and is in good agreement with theoretical results. {This implies that the part of the SNR population below the simulated tracks (e.g.~SNRs with small diameter) makes the slope steeper than the slopes from theoretical considerations. Thus, this part of the sample is likely to cause the resulting discrepancy between the empirical and theoretical $\Sigma-D$ relation.}

According to the results of Pavlovi\'c et al.~(2018), the points below the line, selected for the simulation filter (Fig.~1), are likely to be objects with $\Sigma = 10^{-22}$ and have $D <50$ pc. This part of the $\Sigma-D$ plane is probably mostly populated by dynamically evolved objects that have expanded in environments of very high density. They are likely to be in the post Sedov-Taylor, radiative or dissipating phase of their evolution, following evolution trajectories with different slope than the slope for the Sedov-Taylor phase. In addition, their low brightness makes them very hard to separate from the background which makes them susceptible to observation selection effects. For these reasons, this part of the $\Sigma-D$ plane is likely to have a high scatter which gives a very low $\Sigma-D$ correlation. Such objects, biased by sensitivity or other selection effects, should not be considered as representative of the SNR evolution trajectories and considered as calibrators.

The given simulation filter line in Fig.~1 also serves as a guide for comparative purposes of slopes of different filters against the theoretically derived slope of $-5$. The slopes of filters (ii) and (v) are approximately parallel to this line. The orthogonal fit line for the filter (ii) is translated to the $+\log D$ side when compared to the filter (v), because the filter (ii) cuts out the data points in the lower left part of the $\log \Sigma-\log D$ plane. The filter (v) might thus be more suitable for distance calibration of empirical samples that have low sensitivity related biases. The filter (v) has the largest average fractional error ($f$), almost two times larger than filter (ii) which has the smallest $f$ of all filters. In order to obtain distances with smallest uncertainties, the simulation filter should be preferred for objects that are above the simulation filter cut line.

\vskip.5cm \noindent
\parbox{\columnwidth}{
{\bf Table 2.} Parameters of considered filters. Rows (top to bottom): name of the filter, number of calibration SNRs, smoothing bandwidth in $\log D$, smoothing bandwidth in $\log \Sigma$, value of $\log \Sigma-\log D$ fit intersection with the $\log \Sigma$ axis, value of $\log \Sigma-\log D$ fit slope, value of $\log \Sigma-\log D$ fit slope (angle with the -$\log D$ axis, clockwise), distance fractional error for the fit line, distance fractional error for the mean values of data density distribution in  $\log D - \log \Sigma $  (mean value of $\log D$ distribution for the given value of $\log \Sigma$), same as previous but for the mode value, same as previous but for the median value, confidence interval determined by the distance of the sample from the median value of the slopes of randomly selected re-samples of the same size without repetition of the selected points within the generated re-sample (Fig.~2).\\
\centerline{\begin{tabular}{|l|| @{\hspace{3mm}} c @{\hspace{3mm}} c @{\hspace{3mm}} c @{\hspace{3mm}} c @{\hspace{3mm}} c @{\hspace{3mm}} c|}
\hline
Filter&None&Simulation&Poor dist.&Free exp.&Mixed morph.&Outliers \\
\hline
  $N$  & $ 110$  & $ 68$  & $ 91$  & $ 107$  & $ 77$  & $ 107$ \\
 $h_{\log D}$  & $0.18 $  & $0.14 $  & $0.17 $  & $0.19 $  & $0.19 $  & $0.15 $ \\
 $h_{\log\Sigma}$  & $0.50 $  & $0.48 $  & $0.52 $  & $0.34 $  & $0.50 $  & $0.38 $ \\
 $\log A$  & $-12 \pm 2$  & $-13 \pm 2$  & $-12 \pm 3$  & $-10 \pm 11$  & $-14 \pm 2$  & $-11 \pm 3$ \\
 $\beta$  & $-6 \pm 2$  & $-4.8 \pm 0.9$  & $-6 \pm 2$  & $-7 \pm 8$  & $-5 \pm 2$  & $-7 \pm 3$ \\
 $\beta[^{\circ}]$  & $-80 \pm 2$  & $-78 \pm 2$  & $-80 \pm 2$  & $-82 \pm 3$  & $-79 \pm 3$  & $-81 \pm 2$ \\
 $f_\beta$  & $66.03 \% $  & $36.44 \% $  & $65.28 \% $  & $65.66 \% $  & $69.86 \% $  & $64.01 \% $ \\
 $f_\mathrm{mean}$  & $63.69 \% $  & $31.96 \% $  & $62.34 \% $  & $62.82 \% $  & $66.67 \% $  & $60.45 \% $ \\
 $f_\mathrm{mode}$  & $73.85 \% $  & $29.71 \% $  & $78.46 \% $  & $69.33 \% $  & $76.78 \% $  & $72.06 \% $ \\
 $f_\mathrm{median}$  & $64.56 \% $  & $30.82 \% $  & $64.23 \% $  & $63.85 \% $  & $66.52 \% $  & $62.39 \% $ \\
 conf.~int.~[$\%$]  & $59.86$  & $88.04$  & $6.63$  & $99.98$  & $89.98$  & $99.48$ \\
\hline
\end{tabular}}
\label{tab:filter_score}
} \vskip.5cm

When excluding objects that are in free expansion or in the transition from free expansion to Sedov-Taylor phase of evolution (filter iv), the obtained slope is steeper  than for the full sample, which is in contradiction with theoretical studies (Pavlovi\'c et al.~2018, Berezhko and V{\"o}lk 2004).
Filter (vi) gives similar results. The steep slope of $-7$ in both cases does not appeal to theoretical findings. Such a slope might also be a statistical ``fluke'' because of the exclusion of Cas A source (the brightest point in the sample) located in the upper left part of the $\log \Sigma-\log D$ plane and the tendency of slope to change significantly with small change of the angle between the fit line and the horizontal axis, when steep slopes are concerned (see Section 2.5). 

Excluding objects with poorly determined distances does not make a significant difference in slope, as well as other parameters, when compared to the full sample. Even the fractional error, which is indicative of the data spread around the fit line, is similar. While these objects do appear to be scattered across the area subtended by the whole sample, their removal does not make the slope shallower. On the other hand, the shape of the contour lines points to a significant scatter and absence of correlation in all samples from Fig.~1. Apart from the upper left part of the sample (small objects of high brightness in late free expansion or early Sedov-Taylor phase), the contour lines, especially the central ones, appear to be vertically aligned, except the very small slant for the simulation filter. This implies that the ``true'' slope for the middle and low brightness region of the sample might be influenced with scatter caused by the overall lack of accurate distance estimates.

The results of Pavlovi\'c et al.~(2018) point out that objects in environments of higher densities, populating low brightness and small diameter part of the $\log \Sigma-\log D$ plane, might be the cause for steeper empirical slope. Future studies with specialized hydrodynamical simulations (Kosti\'c 2019) and with better assessed environmental densities of observed SNRs will give more clues in this direction.   

In Fig.~2, the confidence interval map is presented for the considered sample. This should give an insight into the specific behaviour of the sample in terms of sensitivity to the exclusion of data points, i.e., estimating the filter performance against the number of data points in the filtered sample. The filters that are close to the median line, are likely to be of no statistical significance because the similar slope could be achieved by excluding the same number of randomly selected points. Thus, the filters that give slopes further away from the median values, are more distinctive and influential (e.g., they change the geometry of the sample). Since all the filters give statistically similar slopes, this way the impact of the filters can be stratified. 

It is evident that four filters that change the slope by $1$ are at the far end from the median value at $\approx 90-100\%$ confidence limits. The mixed-morphology filter is more appealing from the physical point of view since it is empirical in nature and excludes objects based on morphological criteria, while the simulation based filter is based more on theoretical considerations. The obvious change in the geometry of the sample, upon the application of the simulation filter cut, is similar to the change obtained with the MM filter (both are at $\approx90\%$ confidence limits).

One should bear in mind that mixed-morphology SNRs do not actually follow classical SNR evolution and so can hardly be represented by the Sedov-Taylor model. Even shell SNRs that tend to expand in the high density environments may become ``evolutionary old'' while still of relatively small diameter in comparison to those shell SNRs that expand in the low density medium. Furthermore, in some SNRs from our sample (like Cygnus Loop) both radiative and non-radiative shocks are detected, so a complex evolution must be taking place.

The analysis presented so far points that objects with poorly determined distances might not be the sole or primary  cause of poor $\Sigma-D$ calibration. The objects that have mixed-morphology nature or generally evolve in dense environments, are a more likely cause of scatter in the $\Sigma-D$ data, causing larger fractional errors in distance determination. While the simulation filter has fractional errors of $\approx 30\%$, the rest of the filters have more than double of that value, $\approx 60-75\%$. The simulation filter dismisses $\approx 40\%$ of objects, but the resulting fractional errors are, on average, more than $50\%$ smaller then for other filters. This highlights the low brightness and small diameter objects, part of the sample below the simulation filter cut line, as part of the data with higher scatter when compared with the SNRs of higher brightness. This is likely because of the sensitivity related selection effects of the corresponding surveys and overall difficulties in estimating the diameters and flux densities of small or low brightness objects.    

Considering the fractional errors from the kernel smoothed contours (Table 2), the simulation based filter has similar values for all three parameters (mean, mode and median), while mixed-morphology filter has the smallest fractional error against median values. For consistency, we will use median values in both cases for the purpose of $\Sigma-D$ calibration and the calculation of distances to newly discovered SNRs in the next section (Table 3).

\section{Distances to newly discovered SNRs}

In Table 3 we present several recently discovered SNRs from the literature, which do not have independently derived distances. We calculate distances to these SNRs using $\Sigma-D$ relation. According to the presented analysis in the previous section, the distances are calculated from the orthogonal fit lines and according to median values of the kernel smoothed contours for the full sample, mixed-morphology and simulation based filters. We recalculate distances to $5$ newly discovered SNRs from Gerbrandt et al.~(2014), which were also given in the Table 3 of Paper II. Additionally we include $27$ new candidates for shell SNRs, which were discovered by The \hbox{H\,{\sc i}}, OH, Recombination line survey of the Milky Way -- THOR (Anderson et al.~2017). 

The results from Table 3 point out the importance of the parameter $\log A$. Although the simulation and mixed-morphology filter give similar slope values, the difference in calculated distances for newly discovered remnants is significant because of the different values for $\log A$. Smaller value for $\log A$ in the case of the mixed-morphology filter gives systematically smaller distances when compared to the simulation filter. Also, the slope difference between the full sample and the mixed-morphology filter does not appear to make a significant difference in the calculated distances. The distances from the full sample are slightly larger when compared to mixed-morphology filter, but still noticeably smaller than for the simulation filter. This can be explained with all of the new objects having surface brightness smaller than $10^{-19}\,\mathrm{Wm^{-2}Hz^{-1}sr^{-1}}$ and the similar position of the orthogonal fit lines in the low brightness part of the plot for the MM filter and the full sample.   

As discussed in Section 3, for the purpose of the evolution studies of SNRs, it makes sense to use shallower slopes than the one obtained from the full sample. However, when distance determination is concerned, there is a very small discrepancy between the mixed-morphology filter and the full sample, despite their slope difference. The results of this work suggest that future calibrations and theoretical models could strongly benefit from better insights into SNRs that evolve in dense environments and have low brightness, rather than the overall improvements of distance determination methods.

\vskip.5cm \noindent
\parbox{\columnwidth}{\footnotesize
{\bf Table 3.} Distances to newly discovered SNRs, calculated using $\Sigma-D$ calibrations from this paper. First $5$ SNRs are from Gerbrandt et al.~(2014), and the others from Anderson et al.~(2017). The newly estimated distances are given in the last six columns in the following order: full sample (orthogonal fit distance and distance from the median values of the smoothed data distribution), simulation filter (same as previous filter), mixed-morphology filter (same as previous filter).\\
\centerline{\begin{tabular}{|c|c|c|c|c|c|c|c|c|c|c|}
\hline  
\textbf{No.} &   \textbf{Catalog name}    &   \textbf{Flux dens.}     &     \textbf{Ang. size}            & \textbf{Surf. brightness} &
\multicolumn{2}{c|}{\underline{\bf Full sample}} & \multicolumn{2}{c|}{\underline{\bf Simulation}} & \multicolumn{2}{c|}{\underline {\bf Mixed morph.}} \\ 
  &       &   $S_{1\mathrm{GHz}} $  & $\theta$ &  $\log \Sigma$  &{$d_\mathrm{orth}$}&{$d_\mathrm{median}$} &
{$d_\mathrm{orth}$}& $d_\mathrm{median}$ &
{$d_\mathrm{orth}$}& $d_\mathrm{median}$ \\ 
              &                   &       \textbf{$ (\mathrm{Jy})$}         &  \textbf{$\mathrm{(arcmin)}$} & \textbf{$(\mathrm{Wm^{-2}Hz^{-1}sr^{-1}})$} & \textbf{$\mathrm{(kpc)}$}  & \textbf{$\mathrm{(kpc)}$}    &
              \textbf{$\mathrm{(kpc)}$}  & \textbf{$\mathrm{(kpc)}$}    &
             \textbf{$\mathrm{(kpc)}$}  &  \textbf{$\mathrm{(kpc)}$}\\
\hline
\hline
   1 &    G108.5+11.0    &  0.734    &   $64.9\times39.0$     & -19.36 & 1.08 &  1.47 & 1.54 &  2.09 & 0.93 &  1.31 \\
  2 &   G128.5+2.6      &  0.255    &   $39.6\times21.5$     & -19.35 & 1.85 &  2.51 & 2.63 &  3.60 & 1.59 &  2.22 \\
  3 &   G149.5+3.2      &  0.590    &   $55.6\times49.3$     & -19.49 & 1.09 &  1.49 & 1.57 &  2.08 & 0.95 &  1.38 \\
  4 &    G150.8+3.8     &  0.665    &   $64.1\times18.8$    & -19.08 & 1.40 &  1.71 & 1.95 &  2.71 & 1.18 &  1.22 \\
  5 &  G160.1-1.1       &  0.265    &   $35.9\times13.2$     & -19.07 & 2.23 &  2.69 & 3.10 &  4.29 & 1.88 &  1.92 \\
  6 &    G17.80-0.02    &  0.343    &   $8.8\times8.8$     & -21.18 & 12.54 &  10.69 & 20.96 &  16.40 & 12.40 &  10.65 \\
  7 &  G18.45-0.42      &  2.556    &   $15.2\times15.2$      & -20.78 & 6.22 &  5.96 & 10.03 &  8.80 & 5.96 &  5.62 \\
  8 &   G18.53-0.8      &  0.509    &   $17.2\times17.2$       & -21.59 & 7.53 &  6.02 & 13.06 &  9.87 & 7.69 &  6.39 \\
  9 &   G20.30-0.06     &  0.225    &   $6.2\times6.2$        & -21.06 & 16.98 &  14.96 & 28.08 &  22.56 & 16.64 &  14.60 \\
  10 &    G22.32+0.1    &  1.016    &   $11\times11$        & -20.90 & 9.00 &  8.29 & 14.67 &  12.37 & 8.71 &  7.92 \\
  11 &    G23.85-0.18   &  0.402    &   $5.4\times5.4$        & -20.68 & 16.85 &  16.65 & 26.96 &  24.60 & 16.05 &  15.59 \\
  12 &  G25.49+0.0      &  2.591    &   $14.8\times14.8$       & -20.75 & 6.31 &  6.12 & 10.16 &  9.04 & 6.04 &  5.73 \\
  13 &  G26.13+0.1      &  4.449    &   $22.6\times22.6$       & -20.88 & 4.35 &  4.03 & 7.09 &  6.00 & 4.21 &  3.84 \\
  14 &   G26.53+0.07    &  6.709    &   $22.4\times22.4$        & -20.70 & 4.08 &  4.01 & 6.54 &  5.93 & 3.89 &  3.77 \\
  15 &  G27.06+0.0      &  5.100    &   $15\times15$      & -20.47 & 5.58 &  5.95 & 8.75 &  8.67 & 5.22 &  5.55 \\
  16 & G27.78-0.33      &  0.225    &   $7.4\times7.4$       & -21.21 & 15.10 &  12.80 & 25.32 &  19.70 & 14.98 &  12.93 \\
  17 &   G28.36+0.2     &  2.662    &   $12.8\times12.8$        & -20.61 & 6.92 &  7.02 & 10.99 &  10.31 & 6.55 &  6.55 \\
  18 &   G28.56+0.0     &  1.053    &   $3\times3$        & -19.75 & 21.12 &  27.87 & 31.10 &  38.43 & 18.71 &  26.56 \\
  19 &   G28.64+0.2     &  6.981    &   $22.8\times22.8$    & -20.69 & 4.01 &  3.94 & 6.42 &  5.83 & 3.82 &  3.69 \\
  20 &   G28.78-0.44    &  1.929    &   $13.2\times13.2$       & -20.78 & 7.16 &  6.86 & 11.55 &  10.13 & 6.87 &  6.47 \\
  21 & G28.88+0.4       &  2.331    &   $17.8\times17.8$       & -20.96 & 5.69 &  5.16 & 9.32 &  7.70 & 5.53 &  4.95 \\
  22 &  G29.41-0.18     &  1.278    &   $15\times15$       & -21.07 & 7.05 &  6.18 & 11.68 &  9.36 & 6.92 &  6.06 \\
  23 &  G31.22-0.02     &  0.651    &   $6.2\times6.2$     & -20.59 & 14.18 &  14.50 & 22.50 &  21.20 & 13.41 &  13.49 \\
  24 &  G34.93-0.24     &  0.911    &   $16.2\times16.2$      & -21.28 & 7.10 &  5.93 & 11.98 &  9.22 & 7.08 &  6.05 \\
  25 & G36.66-0.50      &  1.526    &   $16.4\times16.4$  & -21.07 & 6.45 &  5.65 & 10.68 &  8.56 & 6.33 &  5.54 \\
  26 &  G36.68-0.14     &  2.556    &   $20\times20$   & -21.02 & 5.18 &  4.62 & 8.54 &  6.95 & 5.07 &  4.48 \\
  27 & G37.88+0.32      &  3.609    &   $22.8\times22.8$ & -20.98 & 4.48 &  4.03 & 7.37 &  6.05 & 4.37 &  3.89 \\
  28 & G39.56-0.32      &  1.408    &   $17\times17$ & -21.13 & 6.39 &  5.51 & 10.64 &  8.37 & 6.30 &  5.46 \\
  29 & G41.95-0.18      &  1.408    &   $14\times14$  & -20.97 & 7.26 &  6.56 & 11.91 &  9.82 & 7.07 &  6.31 \\
  30 &  G46.54-0.03     &  1.006    &   $12.4\times12.4$  & -21.01 & 8.33 &  7.43 & 13.71 &  11.13 & 8.13 &  7.20 \\
  31 & G47.36-0.09      &  4.236    &   $49.2\times49.2$  & -21.58 & 2.62 &  2.10 & 4.55 &  3.44 & 2.68 &  2.22 \\
  32 & G52.37-0.70      &  6.200    &   $35.4\times35.4$  & -21.13 & 3.06 &  2.65 & 5.09 &  4.02 & 3.01 &  2.62 \\
\hline
\end{tabular}}
}



\section{Summary}

The main results from this paper are summarised as follows:

i) The updated empirical radio $\Sigma-D$ relations based on the new calibration sample (110 SNRs) are presented.

ii) For the first time the kernel density smoothing method is applied for the Galactic SNR sample.

iii) For examined cases, the full sample and each of the filtered samples, the values for angle between the fit line and the $-\log D$ axis overlap within the estimated uncertainties. Thus, the values for angle (slope) should not be used to discriminate between the filters when distance estimates are concerned. Instead, the quality of distance calibrations could be judged from the scatter of the sample, indicated with the fractional error value.

iv) By using our full and filtered samples we estimate distances to 5 newly discovered SNRs and 27 new candidate for shell-like SNRs.

{{\bf Acknowledgements.} We thank an anonymous  referee for the comments that have improved the clarity of the paper. B.V. thanks Petar Kosti\'c for many useful discussions regarding the hydrodynamics of supernova remnants in the non-uniform interstellar medium. The authors acknowledge financial support by the Ministry of Education, Science and Technological Development of the Republic of Serbia through the project 176005 'Emission nebulae: structure and evolution'.}

{\bf  Note added in proof.} While this work was in the proof reading stage, Hurley-Walker et al. (2019ab) reported newly discovered SNRs and SNR candidates. Using the same procedure as for the objects from Table 3 we estimate distances to these objects as well (Table 4).

\vskip.5cm
\noindent 
\parbox{\columnwidth}{\footnotesize
{\bf Table 4.}. Distances to newly discovered SNR objects, calculated using $\Sigma-D$ calibrations from this paper. First $14$ objects are SNRs from Hurley-Walker et al. (2019a), while object number 15 is considered a probable SNR in the same work. The rest of the objects are candidate SNRs from Hurley-Walker et al. (2019b). The table columns have the same meaning as in Table 3.\\
\centerline{\begin{tabular}{|c|c|c|c|c|c|c|c|c|c|c|}
\hline  
\textbf{No.} &   \textbf{Catalog name}    &   \textbf{Flux}     &     \textbf{Ang. size}            & \textbf{Surf.} &
\multicolumn{2}{c|}{\underline{\bf Full sample}} & \multicolumn{2}{c|}{\underline{\bf Simulation}} & \multicolumn{2}{c|}{\underline {\bf Mixed morph.}} \\ 
 &      &   dens.    &              & \textbf{brightness} &
\multicolumn{2}{c|}{} & \multicolumn{2}{c|}{} & \multicolumn{2}{c|}{} \\ 
  &       &   $S_{1\mathrm{GHz}} $  & $\theta$ &  $\log \Sigma$  &{$d_\mathrm{orth}$}&{$d_\mathrm{median}$} &
{$d_\mathrm{orth}$}& $d_\mathrm{median}$ &
{$d_\mathrm{orth}$}& $d_\mathrm{median}$ \\ 
              &                   &       \textbf{$ (\mathrm{Jy})$}         &  \textbf{$\mathrm{(arcmin)}$} & \textbf{$(\mathrm{W/m^{2}/Hz/sr})$} & \textbf{$\mathrm{(kpc)}$}  & \textbf{$\mathrm{(kpc)}$}    &
              \textbf{$\mathrm{(kpc)}$}  & \textbf{$\mathrm{(kpc)}$}    &
             \textbf{$\mathrm{(kpc)}$}  &  \textbf{$\mathrm{(kpc)}$}\\
\hline
\hline
1 & G189.6+3.3  & 15.21 & 90.00 $\times$ 90.00 & -21.55 & 1.42 &  1.14 & 2.45 &  1.85 & 1.44 &  1.20 \\
 2 & G345.1-0.2  & 1.42 & 6.00 $\times$ 6.00 & -20.23 & 12.70 &  14.73 & 19.50 &  21.02 & 11.67 &  13.84 \\
 3 & G345.1+0.2  & 0.64 & 10.00 $\times$ 10.00 & -21.02 & 10.37 &  9.24 & 17.08 &  13.89 & 10.13 &  8.96 \\
 4 & G348.8+1.1  & 0.58 & 10.00 $\times$ 10.00 & -21.06 & 10.54 &  9.27 & 17.44 &  13.99 & 10.33 &  9.05 \\
 5 & G353.3-1.1  & 24.19 & 60.00 $\times$ 60.00 & -21.00 & 1.71 &  1.54 & 2.82 &  2.30 & 1.67 &  1.48 \\
 6 & G359.2-01.1  & 0.46 & 4.00 $\times$ 5.00 & -20.46 & 18.66 &  19.96 & 29.27 &  28.99 & 17.47 &  18.63 \\
 7 & G3.1-0.7 & 4.93 & 28.00 $\times$ 52.00 & -21.29 & 3.03 &  2.52 & 5.11 &  3.93 & 3.02 &  2.58 \\
 8 & G7.5-1.7  & 18.03 & 98.40 $\times$ 98.40 & -21.55 & 1.30 &  1.04 & 2.24 &  1.70 & 1.32 &  1.10 \\
 9 & G13.1-0.5  & 11.43 & 38.00 $\times$ 28.00 & -20.79 & 2.91 &  2.78 & 4.70 &  4.10 & 2.80 &  2.62 \\
 10 & G15.51-0.15  & 1.18 & 8.00 $\times$ 9.00 & -20.61 & 10.42 &  10.60 & 16.55 &  15.55 & 9.86 &  9.85 \\
 11 & MAGPIS 09.6833-0.0667 & 2.78 & 8.50 $\times$ 8.50 & -20.24 & 9.00 &  10.40 & 13.84 &  14.84 & 8.28 &  9.77 \\
 12 & MAGPIS 12.2694+0.2972 & 0.74 & 4.00 $\times$ 4.00 & -20.16 & 18.54 &  21.94 & 28.30 &  31.10 & 16.95 &  20.69 \\
 13 & MAGPIS 28.3750+0.2028 & 1.31 & 10.00 $\times$ 10.00 & -20.71 & 9.18 &  9.05 & 14.72 &  13.28 & 8.76 &  8.48 \\
 14 & MAGPIS 28.7667-0.4250 & 0.89 & 9.50 $\times$ 9.50 & -20.83 & 10.14 &  9.53 & 16.44 &  14.18 & 9.77 &  9.05 \\
 15 & G356.6+00.1  & 0.32 & 7.00 $\times$ 8.00 & -21.06 & 14.10 &  12.39 & 23.32 &  18.69 & 13.82 &  12.10 \\
 16 & G0.1-9.7             & 0.39 & 66.00 $\times$ 66.00 & -22.87 & 3.24 &  1.65 & 6.29 &  3.04 & 3.65 &  1.48 \\
 17 & G2.1+2.7  & 6.63 & 72.00 $\times$ 62.00 & -21.65 & 1.99 &  1.57 & 3.47 &  2.61 & 2.04 &  1.68 \\
 18 & G7.4+0.3  & 0.63 & 18.00 $\times$ 14.00 & -21.42 & 7.66 &  6.24 & 13.09 &  9.94 & 7.72 &  6.52 \\
 19 & G18.9-1.2  & 1.53 & 68.00 $\times$ 60.00 & -22.25 & 2.62 &  1.80 & 4.83 &  3.22 & 2.82 &  1.80 \\
 20 & G19.1-3.1  & 0.91 & 32.00 $\times$ 32.00 & -21.87 & 4.53 &  3.46 & 8.05 &  5.97 & 4.73 &  3.68 \\
 21 & G19.7−0.7  & 4.76 & 28.00 $\times$ 28.00 & -21.04 & 3.74 &  3.30 & 6.17 &  4.96 & 3.66 &  3.21 \\
 22 & G21.8+0.2  & 13.86 & 64.00 $\times$ 42.00 & -21.11 & 2.07 &  1.80 & 3.45 &  2.73 & 2.04 &  1.77 \\
 23 & G23.1+0.1  & 6.18 & 26.00 $\times$ 26.00 & -20.86 & 3.75 &  3.51 & 6.10 &  5.22 & 3.62 &  3.32 \\
 24 & G24.0-0.3  & 10.11 & 48.00 $\times$ 48.00 & -21.18 & 2.30 &  1.97 & 3.85 &  3.01 & 2.28 &  1.97 \\
 25 & G25.3-1.8 & 8.24 & 76.00 $\times$ 94.00 & -21.76 & 1.64 &  1.28 & 2.89 &  2.17 & 1.70 &  1.37 \\
 26 & G28.3+0.2  & 1.36 & 14.00 $\times$ 14.00 & -20.98 & 7.30 &  6.56 & 12.00 &  9.85 & 7.12 &  6.33 \\
 27 & G28.7-0.4  & 1.63 & 10.00 $\times$ 10.00 & -20.61 & 8.85 &  8.99 & 14.06 &  13.19 & 8.38 &  8.39 \\
 28 & G35.3-0.0  & 6.89 & 26.00 $\times$ 22.00 & -20.74 & 3.89 &  3.79 & 6.26 &  5.59 & 3.72 &  3.54 \\
 29 & G230.4+1.2 & 1.33 & 54.00 $\times$ 40.00 & -22.03 & 3.32 &  2.45 & 5.98 &  4.30 & 3.51 &  2.55 \\
 30 & G232.1+2.0  & 2.83 & 50.00 $\times$ 76.00 & -21.95 & 2.42 &  1.83 & 4.34 &  3.16 & 2.54 &  1.92 \\
 31 & G349.1-0.8 & 0.97 & 14.00 $\times$ 14.00 & -21.13 & 7.73 &  6.67 & 12.87 &  10.17 & 7.62 &  6.60 \\
 32 & G350.7+0.6 & 11.20 & 56.00 $\times$ 80.00 & -21.42 & 1.82 &  1.48 & 3.10 &  2.36 & 1.83 &  1.55 \\
 33 & G350.8+5.0 & 10.68 & 72.00 $\times$ 52.00 & -21.37 & 1.94 &  1.60 & 3.30 &  2.52 & 1.95 &  1.65 \\
 34 & G351.0-0.6 & 0.18 & 12.00 $\times$ 12.00 & -21.73 & 11.39 &  8.93 & 20.00 &  15.05 & 11.76 &  9.51 \\
 35 & G351.4+0.4 & 1.70 & 9.00 $\times$ 9.00 & -20.50 & 9.42 &  9.92 & 14.82 &  14.51 & 8.84 &  9.26 \\
 36 & G351.4+0.2 & 0.42 & 18.00 $\times$ 14.00 & -21.60 & 8.20 &  6.52 & 14.24 &  10.77 & 8.39 &  6.97 \\
 37 & G351.9+0.1  & 0.91 & 20.00 $\times$ 16.00 & -21.37 & 6.65 &  5.48 & 11.31 &  8.64 & 6.68 &  5.67 \\
 38 & G353.0+0.8 & 3.30 & 96.00 $\times$ 66.00 & -22.11 & 1.99 &  1.44 & 3.62 &  2.54 & 2.12 &  1.48 \\
 39 & G355.4+2.7 & 0.41 & 22.00 $\times$ 22.00 & -21.89 & 6.64 &  5.06 & 11.83 &  8.74 & 6.94 &  5.35 \\
 40 & G356.5-1.9 & 4.75 & 36.00 $\times$ 48.00 & -21.38 & 2.88 &  2.37 & 4.90 &  3.72 & 2.89 &  2.45 \\
 41 & G358.3-0.7 & 6.02 & 34.00 $\times$ 42.00 & -21.20 & 2.94 &  2.51 & 4.93 &  3.85 & 2.92 &  2.51 \\
  \hline
\end{tabular}}
} 
\vskip.5cm

\newpage
{\bf References\\}\\

Alarie, A., Bilodeau, A. and Drissen, L.: 2014, \journal{Mon. Not. R. Astron. Soc.}, \vol{441}, 2996.

Andersen, M., Rho, J., Reach, W.~T., Hewitt, J.~W. and Bernard, J.~P.: 2011, \journal{Astrophys. J.}, \vol{742}, 7.

Anderson, L.~D., Wang, Y., Bihr, S., Rugel, M., Beuther, H., Bigiel, F., Churchwell, E. et al.: 2017, \journal{Astron. Astrophys.}, \vol{605}, A58.

Arias, M., Dom{\v c}ek, V., Zhou, P. and Vink, J.: 2019, \journal{Astron. Astrophys.}, \vol{627}, A75.

Bamba, A., Yokogawa, J., Sakano, M. and Koyama, K.: 2000, \journal{Publ. Astron. Soc. Jpn.}, \vol{52}, 259.

Berezhko, E. G. and V{\"o}lk, H. J.: 2004, \journal{Astron. Astrophys.}, \vol{427}, 525-536.

Borkowski, K.~J., Reynolds, S.~P. and Roberts, M.~S.~E.: 2016, \journal{Astrophys. J.}, \vol{819}, 160.

Bozzetto, L.~M. et al.: 2017, \journal{Astrophys. J. Suppl. Ser.}, \vol{230(1)}, 2, 30 pp. 

Brisken, W.~F., Carrillo-Barrag\'{a}n, M., Kurtz, S., Finley, J.~P.: 2006, \journal{Astrophys. J.}, \vol{652}, 554.

Camilo, F., Kaspi, V.~M., Lyne, A.~G., Manchester, R.~N., Bell, J.~F., D'Amico, N., McKay, N.~P.~F., Crawford, F.: 2000, \journal{Astrophys. J.}, \vol{541}, 367.

Castelletti, G., Dubner, G., Brogan, C. and Kassim, N.~E.: 2007, \journal{Astron. Astrophys.}, \vol{471}, 537.

Castelletti, G., Giacani, E., Dubner, G., Joshi, B.~C., Rao, A.~P. and Terrier, R.: 2011, \journal{Astron. Astrophys.}, \vol{536}, A98.

Caswell, J.~L., McClure-Griffiths, N.~M. and Cheung, M. C. M.: 2004, \journal{Mon. Not. R. Astron. Soc.}, \vol{352}, 1405. 

Caswell, J.~L., Murray, J.~D., Roger, R.~S., Cole, D.~J. and Cooke, D.~J.: 1975, \journal{Astron. Astrophys.}, \vol{45}, 239.

Chatterjee, S., Brisken, W.~F., Vlemmings, W.~H.~T., Goss, W.~M., Lazio, T.~J.~W., Cordes, J.~M., Thorsett, S.~E.; Fomalont, E.~B., Lyne, A.~G. and Kramer, M.: 2009, \journal{Astrophys. J.}, \vol{698}, 250.

Condon, J. J., Cotton, W. D., Greisen, E. W., Yin, Q. F., Perley, R. A., Taylor,G. B., and  Broderick, J. J.: 1998, \journal{Astron. J.}, \vol{115(5)}, 1693-1716.

Dickel, J.~R., Milne, D.~K. and Strom, R.~G.: 2000, \journal{Astrophys. J.}, \vol{543}, 840.

Doherty, M., Johnston, S., Green, A.~J., Roberts, M.~S.~E., Romani, R.~W., Gaensler, B.~M. and Crawford, F.: 2003, \journal{Mon. Not. R. Astron. Soc.}, \vol{339}, 1048.

Duin, R. P. W.: 1976, \journal{IEEE Transactions on Computers}, \vol{25}, 1175.

Efron, B. and Tibshirani, R.~J.: 1994, An introduction to the bootstrap, CRC press.

Eger, P., Rowell, G., Kawamura, A., Fukui, Y., Rolland, L. and Stegmann, C.: 2011, \journal{Astron. Astrophys.}, \vol{526}, A82.

Ferrand, G. and Safi-Harb, S.: 2012, \journal{Adv. Space Res.}, \vol{49}, 1313.

Fesen, R.~A., Weil, K.~E., Cisneros, I.~A., Blair, W.~P. and Raymond, J.~C.: 2018, \journal{Mon. Not. R. Astron. Soc.}, \vol{481}, 1786.

Foster, T.~J., Cooper, B., Reich, W., Kothes, R. and West, J.: 2013, \journal{Astron. Astrophys.}, \vol{549}, A107. 

Foster, T. and Routledge, D.: 2003, \journal{Astrophys. J.}, \vol{598}, 1005.

Foster, T., Routledge, D. and Kothes, R.: 2004, \journal{Astron. Astrophys.}, \vol{417}, 79.

Frail, D.~A., Goss, W.~M., Reynoso, E.~M., Giacani, E.~B., Green, A.~J. and Otrupcek, R.: 1996, \journal{Astron. J.}, \vol{111}, 1651.

Gaensler, B.~M., Manchester, R.~N. and Green, A.~J.: 1998, \journal{Mon. Not. R. Astron. Soc.}, \vol{296}, 813.

Gerbrandt, S., Foster, T.~J., Kothes, R., Geisb{\"u}sch,J. and Tung, A.: 2014, \journal{Astron.  Astrophys}, \vol{566}, A76.

Giacani, E.~B., Dubner, G., Cappa, C. and Testori, J.: 1998, \journal{Astron. Astrophys. Supp.}, \vol{133}, 61.

Giacani, E.~B., Dubner, G.~M., Green, A.~J., Goss, W.~M. and Gaensler, B.~M.: 2000, \journal{Astron. J.}, \vol{119}, 281.

Giacani, E., Smith, M.~J.~S., Dubner, G. and Loiseau, N.: 2011, \journal{Astron. Astrophys.}, \vol{531}, A138.

Giacani, E., Smith, M.~J.~S., Dubner, G., Loiseau, N., Castelletti, G., Paron, S.: 2009, \journal{Astron. Astrophys.}, \vol{507}, 841.

Green, D.~A.: 1984, \journal{Mon. Not. R. Astron. Soc.}, \vol{209}, 449.

Green D.~A.: 2019,\journal{J. Astrophys. Astron.}, \vol{40}, 36. 

Halpern, J.~P., Gotthelf, E.~V. and Camilo, F.: 2012, \journal{Astrophys. J. Lett.}, \vol{753}, L14.

Hewitt, J.~W. and Yusef-Zadeh, F.: 2009, \journal{Astrophys. J. Lett.}, \vol{694}, L16.

Hurley-Walker, N. et al.: 2019a, \journal {PASA}, {\bf 36}, e048.

Hurley-Walker, N. et al.: 2019b, \journal {PASA}, {\bf 36}, e045.

Jackson, M.~S., Safi-Harb, S., Kothes, R. and Foster, T.: 2008, \journal{Astrophys. J.}, \vol{674}, 936.

Jiang, B., Chen, Y. and Wang, Q. D.: 2007, \journal{Astrophys. J.}, \vol{670}, 1142.

Kamitsukasa, F., Koyama, K., Nakajima, H., Hayashida, K., Mori, K., Katsuda, S., Uchida, H. and Tsunemi, H.: 2016, \journal{Publ. Astron. Soc. Jpn.}, \vol{68}, S7.

Katsuda, S., Tanaka, M., Morokuma, T., Fesen, R. and Milisavljevic, D.: 2016, \journal{Astrophys. J.}, \vol{826}, 108.

Katsuda, S., Tsunemi, H. and Mori, K.: 2008, \journal{Astrphys. J. Lett.}, \vol{378}, L35.

Keles, T.: 2018, \journal{International Online Journal of Educational Sciences}, \vol{10(3)}, 200-214.

Kosti\'c, P.: 2019, \journal{Serb. Astron. J.}, submitted.

Kothes, R. and Dougherty, S.~M.: 2007, \journal{Astron. Astrophys.}, \vol{468}, 993.

Kothes, R., Uyanıker, B. and Reid, R.~I.: 2005, \journal{Astron. Astrophys.}, \vol{444}, 871.

Leahy, D.~A.: 2004, \journal{AJ}, \vol{127}, 2277.

Leahy D.~A and Green, K.~S.: 2012, \journal{Astrophys. J.}, \vol{760}, 25.

Leahy, D.~A, and Tian, W.~W.: 2006, \journal{Astron. Astrophys.}, \vol{451}, 251.

Leahy, D.~A, and Tian, W.~W.: 2007, \journal{Astron. Astrophys.}, \vol{461}, 1013.

Lemiere, A., Slane, P., Gaensler, B.~M. and Murray, S.: 2009, \journal{Astrophys. J.}, \vol{706}, 1269.

Lozinskaya, T.~A., Sitnik, T.~G. and Pravdikova, V.~V.: 1993, \journal{Astron. Rep.}, \vol{37}, 240.

Maggi, P., et al.: 2019, \journal{Astron. Astrophys.}, submitted.

Matthews, B.~C., Wallace, B.~J. and Taylor, A.~R.: 1998, \journal{Astrophys. J.}, \vol{493}, 312.

McClure-Griffiths, N.~M., Green, A.~J., Dickey, J.~M., Gaensler, B.~M., Haynes, R.~F. and Wieringa, M.~H.: 2001, \journal{Astrophys. J.}, \vol{551}, 394.

Minter, A.~H., Camilo, F., Ransom, S.~M. Halpern, J.~P. and Zimmerman, N.: 2008, \journal{Astrophys. J.}, \vol{676}, 1189.

Moriguchi, Y., Tamura, K., Tawara, Y., Sasago, H., Yamaoka, K., Onishi, T., Fukui, Y.: 2005, \journal{Astrophys. J.}, \vol{631}, 947.

Nikoli\'c, S., van de Ven, G., Heng, K., Kupko, D., Husemann, B., Raymond, J.~C., Hughes, J.~P. and Falc\'on-Barroso, J.: 2013, \journal{Science}, \vol{340}, 45.

Pavlovi\'c, M.~Z., Uro\v sevi\'c, D., Vukoti\'c, B., Arbutina, B. and G{\"o}ker, {\"U}.~D.:  2013, \journal{Astrophys. J. Suppl. Ser.}, \vol{204}, 4. (Paper I)

Pavlovi\'c, M.~Z., Dobard\v zi\'c, A., Vukoti\'c, B. and Uro\v sevi\'c, D.: 2014, \journal{Serb. Astron. J.}, \vol{189}, 25. (Paper II)

Pavlovi\'c, M.~Z., Uro{\v s}evi\'c, D., Arbutina, B., Orlando, S., Maxted, N. and Filipovi\'c, M.~ D.: 2018, \journal{Astrophys. J.}, \vol{852(2)}, 84.

Pineault, S., Landecker, T.~L., Madore, B. and Gaumont-Guay, S.: 1993, \journal{Astron. J.}, \vol{105}, 1060. 

Prinz, T. and Becker, W.: 2013, \journal{Astron. Astrophys.}, \vol{550}, A33.

Ranasinghe, S. and Leahy, D.~A.: 2017, \journal{Astrophys. J.}, \vol{843}, 119.

Ranasinghe, S. and Leahy, D.~A.: 2018a, \journal{Astron. J.}, \vol{155}, 204.

Ranasinghe, S. and Leahy, D.~A.: 2018b, \journal{Mon. Not. R. Astron. Soc.}, \vol{477}, 2243.

Ranasinghe, S., Leahy, D.~A. and Tian, W.: 2018, \journal{Open Phys. J.}, \vol{4}, 1.

Reynoso, E.~M., Cichowolski, S. and Walsh, A.~J.: 2017, \journal{Mon. Not. R. Astron. Soc.}, \vol{464}, 3029.

Reynoso, E.~M., Johnston, S., Green, A.~J. and Koribalski, B.~S.: 2006, \journal{Mon. Not. R. Astron. Soc.}, \vol{369}, 416.

Rho, J., and Petre, R.: 1998, \journal{Astrphys. J. Lett.}, \vol{503}, 167.

Rosado, M., Ambrocio-Cruz, P., Le Coarer, E. and Marcelin, M.: 1996, \journal{Astron. Astrophys.}, \vol{315}, 243.

Ro\.{z}ko, K., Rajwade, K.~M., Lewandowski, W., Basu, R., Kijak, J., Lorimer, D.~R.: 2018, \journal{Mon. Not. R. Astron. Soc.}, \vol{479}, 2193. 

S\'anchez-Cruces, M., Rosado, M., Fuentes-Carrera, I. and Ambrocio-Cruz, P.: 2018, \journal{Mon. Not. R. Astron. Soc.}, \vol{473}, 1705.

Saito M., Matsumoto M.: 2008, in Monte Carlo and QuasiMonte Carlo Methods 2006, Keller A., Heinrich S., Niederreiter H. (eds.), Springer, Berlin Heidelberg, pp. 607--622.

Sankrit, R., Raymond, J.~C., Blair, W.~P., Long, K.~S., Williams, B.~J., Borkowski, K.~J., Patnaude, D.~J. and Reynolds, S.~P.: 2016, \journal{Astrophys. J.}, \vol{817}, 36.

Sawada, M., Tachibana, K., Uchida, H., Ito, Y., Matsumura, H., Bamba, A., Tsuru, T.~G. and Tanaka, T.: 2019, \journal{Publ. Astron. Soc. Jpn.}, \vol{71}, 61.

Shan, S.~S., Zhu, H., Tian, W.~W., Zhang, M.~F., Zhang, H.~Y., Wu, D. and Yang, A.~Y.: 2018, \journal{Astrophys.  J. Suppl. Ser.}, \vol{238}, 35.

Shan, S.~S, Zhu, H., Tian, W.~W., Zhang, H.~Y., Yang, A.~Y. and Zhang, M.~F.: 2019, \journal{Research in Astronomy and Astrophysics}, \vol{19}, 92S.

Shklovsky, I.~S.: 1960, \journal{Astronomicheskii Zhurnal}, \vol{37(2)}, 256.

Su, H., Tian, W., Zhu, H. and Xiang, F.~Y.: 2014, \journal{	
	Supernova Environmental Impacts, Proceedings of the International Astronomical Union, IAU Symposium}, \vol{296}, 372.

Su, H.-Q., Zhang, M.-F., Zhu, H. and Wu, D.: 2017a, \journal{Res. Astron. Astrophys.}, \vol{17}, 109.

Su, Y., Zhou, X., Yang, J., Chen, X., Chen, Y., Liu, Y., Wang, H., Li, C. and Zhang, S.: 2017b, \journal{Astrophys. J.}, \vol{836}, 211.

Supan, L., Castelletti, G., Supanitsky, A.~D., Burton, M.~G.: 2018, \journal{Astron. Astrophys.}, \vol{619}, A108.

Takata, A., Nobukawa, M., Uchida, H., Tsuru, T.~Go, Tanaka, T., Koyama, K.: 2016, \journal{Publ. Astron. Soc. Jpn.}, \vol{68}, S3.

Tian, W.~W., Haverkorn, M. and Zhang, H.~Y.: 2007, \journal{Mon. Not. R. Astron. Soc.}, \vol{378}, 1283.

Tian, W.~W. and Leahy, D.~A.: 2006, \journal{Astron. Astrophys.}, \vol{455}, 1053.

Tian, W.~W. and Leahy, D.~A.: 2012, \journal{Mon. Not. R. Astron. Soc.}, \vol{421}, 2593.

Tian, W.~W. and Leahy, D.~A.: 2014, \journal{Astrophys. J. Lett.}, \vol{783}, L2.

Tian, W.~W., Leahy, D.~A. and Foster, T.~J.: 2007, \journal{Astron. Astrophys.}, \vol{465}, 907.

Tian, W.~W., Li, Z., Leahy, D.~A., Yang, J., Yang, X.~J., Yamazaki, R. and Lu, D.: 2010, \journal{Astrophys. J.}, \vol{712}, 790.

Uro{\v s}evi{\'c}, D.: 2002, \journal{Serb. Astron. J.}, \vol{165}, 27.

Uro{\v s}evi{\'c}, D.: 2003, \journal{Astrophys. Space. Sci.}, \vol{283}, 75.

Uro{\v s}evi{\'c}, D.: 2005, \journal{Publ. Astron. Soc. "Rudjer Bo{\v s}kovi{\' c}"}, \vol{5}, 113.

Uro{\v s}evi{\'c}, D., Pannuti, T. G., Duric, N. and Theodorou, A.: 2005, \journal{Astron. Astrrophys}, \vol{435}, 437.

Uro\v sevi\'c, D., Vukoti\'c, B., Arbutina, B. and Sarevska, M.: 2010, \journal{Astrophys. J.}, \vol{719}, 950.

Uyaniker, B., Kothes, R. and Brunt, C.~M.: 2002, \journal{Astrophys. J.}, \vol{565}, 1022.

Vel\'azquez, P.~F., Dubner, G.~M., Goss, W.~M. and Green, A.~J.: 2002, \journal{Astron. J.}, \vol{124}, 2145.

Vink J.: 2004, \journal{Astrophys. J.}, \vol{604}, 693.

Vukoti\'c, B., Jurkovi\'c, M., Uro\v sevi\'c, D. and Arbutina, B.: 2014, \journal{Mon. Not. R. Astron. Soc.}, \vol{440}, 2026.

Yamauchi, S., Nobukawa, M., Koyama, K. and Yonemori, M.: 2013, \journal{Publ. Astron. Soc. Jpn.}, \vol{65}, 6.

Yamaguchi, H., Ueno, M., Koyama, K., Bamba, A. and Yamauchi, S.: 2004, \journal{Publ. Astron. Soc. Jpn.}, \vol{56}, 1059.

Yar-Uyaniker, A., Uyaniker, B. and Kothes, R.: 2004, \journal{Astrophys. J.}, \vol{616}, 247.

Yu, B., Chen, B.~Q., Jiang, B.~W. and Zijlstra, A.: 2019, , \journal{Mon. Not. R. Astron. Soc.}, \vol{488}, 3129.

Zhang, X., Chen, Y., Li, H. and Zhou, X.: 2013, \journal{Mon. Not. R. Astron. Soc.}, \vol{429}, L25. 

Zhang, G.-Y., Chen, Y., Su, Y., Zhou, X., Pannuti, T.~G. and Zhou, P.: 2015, \journal{Astrophys. J.}, \vol{799}, 103.

Zhang, G.-Y., Slavin, J.~D., Foster, A., Smith, R.~K., ZuHone, J.~A., Zhou, P., and Hen, Y.: 2019, \journal{Astrophys. J.}, \vol{875}, 81.

Zhu, H., Tian, W.~W. and Wu, D.: 2015, \journal{Mon. Not. R. Astron. Soc.}, \vol{452}, 3470.

Zhou, X., Yang, J., Fang, Mi., Su, Y., Sun, Y. and Chen, Y.: 2016, \journal{Astrophys. J.}, \vol{833}, 4.

\end{document}